\documentclass[a4paper,fleqn,usenatbib]{mnras}

\usepackage[T1]{fontenc}
\usepackage{ae,aecompl}
\usepackage{xcolor}
\usepackage[normalem]{ulem}		

\usepackage{graphicx}	 
\usepackage{amsmath}     
\usepackage{amssymb}  
\usepackage[bottom]{footmisc}
\usepackage[varg]{txfonts}
\usepackage{hyperref}

\usepackage{comment}
\usepackage[normalem]{ulem}

\definecolor{myGreen}{rgb}{0.0, 0.5, 0.0}

\title[{Synthetic Spectra from PIC Simulations}]{Synthetic Spectra from Particle-in-cell Simulations of\\
       Relativistic Jets containing an initial Toroidal Magnetic Field}

\author[Du\c{t}an, I. et al.]{Ioana Du\c{t}an,$^{1}$\thanks{E-mail: ioana.dutan@spacescience.ro}
Kenichi Nishikawa,$^{2}$
Athina Meli,$^{3,4}$
Oleh Kobzar,$^{5}$
Christoph K\"ohn,$^{6}$
\newauthor Yosuke Mizuno,$^{7,8}$
Nicholas MacDonald,$^{9}$
Jos\'e L. G\'omez,$^{10}$ and
Kouichi Hirotani$^{11}$
\\
$^{1}$Institute of Space Science -- INFLPR subsidiary, Atomi\c{s}tilor 409, RO-077125 M\u{a}gurele, Romania \\
$^{2}$Department of Physics, Chemistry and Mathematics, Alabama A\&M University, Normal, AL 35762, USA\\
$^{3}$College of Science and Technology, North Carolina A\&T State University, North Carolina, NC 27411, USA\\
$^{4}$STAR Institute, Universite de Liege, Sart Tilman, 4000 Li{\'{e}}ge, Belgium\\
$^{5}$Astronomical Observatory, Jagiellonian University, ul. Orla 171, 30-244 Krak{\'o}w, Poland\\
$^{6}$Technical University of Denmark, National Space Institute (DTU Space), Elektrovej 328, 2800 Kgs Lyngby, Denmark\\
$^{7}$Institute for Theoretical Physics, Goethe University, D-60438 Frankfurt am Main, Germany\\
$^{8}$Tsung-Dao Lee Institute, Shanghai Jiao Tong University, Shanghai, 200240, China \\
$^{9}$Department of Physics and Astronomy, University of Mississippi, 108 Lewis Hall, University, MS 38677, USA\\
$^{10}$Instituto de Astrof\'isica de Andaluc\'ia, CSIC, Apartado 3004, 18080 Granada, Spain\\
$^{11}$Taiwan Institute of Astronomy and Astrophysics, Academia Sinica, Taipei 10617, Taiwan, Republic of China\\
}

\date{Accepted  Received ; in original form }

\pubyear{2024}

\usepackage{hyperref}

\begin{document}
\label{firstpage}
\pagerange{\pageref{firstpage}--\pageref{lastpage}}
\maketitle

\begin{abstract}
The properties of relativistic jets, their interaction with the environment, and their emission of radiation can be self-consistently studied by using collisionless particle-in-cell (PIC) numerical simulations. Using three-dimensional (3D), relativistic PIC simulations, we present the first self-consistently calculated synthetic spectra of head-on and off-axis emission from electrons accelerated in cylindrical, relativistic plasma jets containing an initial toroidal magnetic field. The jet particles are initially accelerated during the linear stage of growing plasma instabilities, which are the Weibel instability (WI), kinetic Kelvin-Helmholtz instability (kKHI), and mushroom instability (MI). In the nonlinear stage, these instabilities are dissipated and generate turbulent magnetic fields which accelerate particles further. We calculate the synthetic spectra by tracing a large number of jet electrons in the nonlinear stage, near the jet head where the magnetic fields are turbulent. Our results show the basic properties of jitter-like radiation emitted by relativistic electrons when they travel through a magnetized plasma with the plasma waves driven by kinetic instabilities (WI, kKHI, and MI) growing into the nonlinear regime. At low frequencies, the slope of the spectrum is $\sim 0.94$, which is similar to that of the jitter radiation, rather than that of the classical synchrotron radiation which is $\sim 1/3$. Although, we start with a weak magnetized plasma, the plasma magnetization increases locally in regions where the magnetic field becomes stronger due to kinetic instabilities. The results of this study may be relevant for probing photon emission from low energies up to, at least, low energies in the X-ray domain in AGN/blazar and GRB jets, as the peak frequency of synthetic spectra increases as the Lorentz factor of the jet increases from 15 to 100.

\end{abstract}



\section{Introduction}\label{intro}

Relativistic jets are observed across the entire electromagnetic spectrum in a variety of astrophysical systems, such as active galactic nuclei \cite[AGNs, e.g.,][]{blandford19}, gamma-ray bursts \cite[GRBs, e.g.,][]{meszaros14}, and some black hole X-ray binaries \cite[e.g.,][]{saikia19}, being launched from compact objects (black holes or neutron stars) typically surrounded by accretion disks. They are well-collimated outflows of plasma and fields, propagating with velocities close to the speed of light \cite[e.g.,][]{lister09}. Hence, relativistic jets are supersonic, producing shocks that lead to a turbulent magnetic field, which can play an important role in accelerating jet particles \cite[e.g.,][]{nishikawa21}. They are also subject to kinetic instabilities driven by currents and shears \cite[e.g.,][]{nishikawa09b} and to magnetic reconnection \cite[e.g.,][]{alves12,meli23}, in three-dimensional (3D) PIC simulations, as well as in current sheets developed within turbulent 3D magnetohydrodynamic (MHD) flows \cite[e.g.,][]{kowal12, deGouveia15,beresnyak16,delValle16} and in instability-driven turbulent relativistic MHD jets \cite[e.g.,][]{davelaar20,medina21,medina23}. Although MHD-PIC simulations \cite[e.g.,][]{medina23} can complement PIC simulations, such simulations cannot explore particle interactions with self-generated electromagnetic fields, nor include kinetic instabilities.  PIC simulations self-consistently solve plasma dynamics, but they are inherently limited by the size and resolution of the plasma system. These PIC simulations are complementary to MHD simulations.

From radio to optical and ultraviolet bands, for many AGN jets, the observed power-law spectra typically correspond to synchrotron emission, which arises from the acceleration of non-thermal energy distributions of particles (electrons/positrons) in a largely ordered magnetic field. Similarly, many GRB afterglows are described reasonably well by synchrotron emission from a highly relativistic structured jet \cite[e.g.,][]{ghirlanda19}. \citet{medvedev99,medvedev00} have proposed an alternative interpretation of the emission in GRB internal and external shocks (afterglow) from relativistic electrons traveling through highly nonuniform, small-scale magnetic fields via the jitter radiation, the spectral power of which scales as $P(\omega)\propto\omega^1$, at low frequencies.

Numerical simulations play a crucial role in the understanding of relativistic jets, including their emission. To provide a better description of the emission from jets, we need to calculate spectra by tracing jet electrons in the time-varying electric and magnetic fields self-consistently, where the magnetic fields and the accelerated particles are produced as part of the plasma evolution. 
PIC methods, with their associated kinetic plasma shocks and instabilities, can self-consistently explain the generation and amplification of magnetic fields, particle acceleration, and emission of radiation in jet plasma \citep{hededal05}. (Here, and throughout this paper, we refer to PIC methods or simulations as being {\it collisionless}.) In PIC simulations individual particles can be followed in their self-generated electromagnetic environment, allowing a better understanding of the statistical properties of wave-particle interactions, where the nonlinear feedback of the scattering processes can be included {\citep{pohl20}}. Since the plasma density in astrophysical jets is very low, i.e., the distance between the particles is larger than the mean-free path of the particles, we use collisionless PIC methods, where the particles interact with the electromagnetic fields created by the particles themselves through kinetic plasma shocks, instabilities, or magnetic reconnection \cite[e.g.,][]{nishikawa14,meli23}. To justify our approach of using PIC methods to describe relativistic jets associated with AGN and GRBs, we provide an estimation for the ratio of the collision frequency to the gyrofrequency, in Section~\ref{simulations}.

Synthetic spectra can be determined by integrating the expression of the radiated power, derived from the Li\'enard--Wiechart potentials \citep{jackson99} for a large number of representative particles in the PIC representation of the plasma (see Eq.~\ref{spect}), e.g., (i) using a test particle approach with a prescribed turbulent magnetic field (as red, white, and blue noises) \citep{hededal05} and (ii) injecting two test jet electrons with different perpendicular velocities in a parallel magnetic field  \citep{nishikawa08, nishikawa09b}. 

In the simulations performed by \citet{sironi09}, a shock is generated by the interaction between a plasma stream reflected from a wall and an incoming plasma stream. The spectra have been obtained from particles accelerated in the shock using electromagnetic fields created near the shock. 

On the one hand, nonthermal radiation spectra in self-consistent simulations can be obtained by directly tracing particles from simulations, as in PIC calculations, without making assumptions about the magnetic field, particle orbit, or other factors while simultaneously solving Maxwell equations. Such self-consistent calculations were performed, by e.g., \citet{frederiksen10} for counter-streaming flows and \citep{nishikawa11,nishikawa12,nishikawa13b} by injecting jets into an ambient medium using a slab model. On the other hand, while magnetohydrodynamics (MHD) is a widely used method for describing the plasma in relativistic jets,  the method cannot perform calculations of radiation spectra by tracing individual particles as the PIC method does. For calculating nonthermal radiation, MHD simulations usually employ ray tracing as a numerical approach. Nevertheless, nonthermal particle acceleration has been demonstrated in MHD simulations. For instance, particles have been injected into several snapshots of a 3D Poynting-flux-dominated MHD jet with moderate magnetization ($\sigma \sim 1$), where current-driven kink instability induces turbulence and rapid magnetic reconnection \citep{medina21}. We note that in such simulations the particles are accelerated by electromagnetic fields generated by MHD, but the kinetic effects are not modeled. However, they can effectively provide the length scale of the jet, comparable to that of an AGN, owing to the scale-free nature of the MHD method. In contrast, PIC simulations are constrained to smaller scales, where the length scale for the propagation of low-frequency electromagnetic radiation in a collisionless plasma is determined by the plasma skin depth \cite[e.g.,][]{macdonald21}. To reach the length scale of an astrophysical jet, PIC simulations must employ exceptionally large simulation grids, which are very challenging to achieve due to the substantial computational resources required. Moreover, using a hybrid approach, \citet{medina23} explore particle acceleration in turbulent MHD relativistic jets transitioning from small to large scales using 3D MHD-PIC simulations and find similar results as in the earlier work of \citet{medina21}. In both MHD and PIC simulations growing instabilities are present, followed by turbulence and magnetic reconnection; however, fully kinetic PIC simulations resolve the microscopic plasma dynamics in great detail. We emphasize that while the instabilities initiating turbulence differ, the resulting acceleration mechanisms share common characteristics, with the primary distinction being the scale at which they are resolved (see also \citet{deGouveia24}).

To provide a more accurate description of PIC plasma jets, injection of a cylindrical plasma jet into an ambient medium is more suitable than using a slab model, since for a simulated jet with a cylindrical geometry the velocity-shear instabilities arise more naturally at the interface between the jet and the ambient plasma. The generation of magnetic fields associated with the velocity shear between an unmagnetized relativistic jet and an unmagnetized sheath plasma via kKHI and MI has previously been studied using a slab model \cite[e.g.,][]{alves12,nishikawa13a,nishikawa14}. 

Moreover, including an initial helical or toroidal magnetic field in the simulated plasma is important as resolved highly collimated relativistic jets exhibit twisted time-dependent structures as indicated by observations \cite[e.g.,][]{lobanov01,pasetto21}.

Therefore, \cite{meli23} have performed PIC simulations for cylindrical relativistic plasma jets containing an initial toroidal magnetic field injected into an ambient plasma at rest with a large simulation grid. They have investigated the growth of WI, kKHI, and MI, simultaneously in the jet plasma deep into the non-linear regime, as well as possible magnetic reconnection sites. 

In this paper, we self-consistently calculate synthetic spectra from electrons accelerated in a cylindrical relativistic jet containing an initial toroidal magnetic field injected into an ambient plasma at rest, starting from the PIC plasma simulation code of \cite{meli23}, where we add subroutines for calculation of radiation. In our simulations, the jet plasma is initially kinetically dominated, with a magnetization parameter $\sigma < 10^{-2}$, and with values that increase due to growing kinetic instabilities during the simulations. Here, we use (i) two different particle species for the jet plasma: electron-positron (e$^{\pm}$) and electron-ion (e$^{-}$- i$^{+}$), (ii) two values for the Lorentz factor of the jet, $\Gamma=15$ and $\Gamma=100$, (iii) two values for the strength of the amplitude of the initially applied toroidal magnetic field, $B_{0}=0.1$ and $B_{0}=0.5$, and (iv) two emission directions, head-on and 5$^\circ$-off. Next, we study the kinetic effects (in the presence of WI, kKHI, and MI) on the generated magnetic field, and consequently on the emission.

The paper is organized as follows. In Sect.~\ref{simulations}, we briefly describe our 
simulation setup and methods. In Sect.~\ref{results}, we present how the global system evolution and electron acceleration occur (Subsect.~\ref{evolution}), the analysis of turbulence via the Fast Fourier Transform (Subsect.~\ref{fft}), and the results on synthetic spectra obtained from PIC simulations for various parameters of the jet plasma (Subsect.~\ref{sec:spectra}). We also discuss their importance for understanding emission of radiation from AGN and GRB jets. Here, the jitter-like spectra are obtained by tracing jet electrons in the nonlinear region where turbulent magnetic fields are generated by dissipated kinetic instabilities which are similar to observed spectra by Fermi \cite[e.g.,][]{abdo2009}. 
Discussions and conclusions follow in Sect.~\ref{conclusions}.

\section{Simulation setup and methods}\label{simulations}

We use TRISTAN-MPI, a highly parallelized, 3D-relativistic PIC simulation code developed by \citet{niemiec08,nishikawa09a,nishikawa14,nishikawa16a,nishikawa16b}, which is based on the improved versions of TRISTAN \citep{buneman93} presented in \citep{nishikawa03,nishikawa05}. The code uses a 2nd order finite difference scheme to advance the equations in time and to calculate spatial derivatives. The electric and magnetic fields are stored on a 3D Yee mesh and an interpolation function is used to interpolate the electric and magnetic fields at the particle positions. These (computational or macro) particles model a large number of real particles through the specific distribution of their charge over the grid cell(s). Here, we use a triangular-shape-cloud approximation, where the shape factor determines the fractions of particle charge deposited at given grid points. (see \citet{nishikawa21} for examples of particle shape factors). The electromagnetic fields are then used to advance the particle velocity under the action of the Lorentz's force (using the Boris' algorithm \citep{boris70}) and the currents are collected. The code has been adapted for injecting a cylindrical, relativistic plasma jet containing an initial toroidal magnetic field into an ambient plasma at rest by \citet{meli23}.

We perform simulations with a grid size of $(L_{x}, L_{y}, L_{z}) = (1285\,\Delta,789\,\Delta,789\,\Delta)$, where $\Delta$ is the size of the grid cell. The size of the numerical grid is large enough to allow for growing instabilities into the nonlinear regime. A plasma jet, having a bulk Lorentz factor of $\Gamma$, is injected in the middle of the $y-z$ plane, $(y_{\rm jc},\, z_{\rm jc}) = (381\,\Delta, \,381\,\Delta)$, at $x = 100\,\Delta$ in order to avoid boundary effects. The jet radius is $r_{\rm jt}=100\,\Delta$, ten times larger than the electron skin depth ($\lambda_{\rm e} =  c/\omega_{\rm pe} = 10\,\Delta$), where $c$ is the speed of light. To increase the numerical stability, we used a time step $\Delta t = 0.1\,\omega_{\rm pe}^{-1}$, where $\omega_{\rm pe}$ is the electron plasma frequency. Some other plasma parameters used in the simulations are the Debye length, $\lambda_{\rm D}=0.5\,\Delta$, the number density of particles of each species per cell for jet, $n_{\rm jt}= 8$, and for ambient, $n_{\rm am} = 12$ (these numbers correspond to billions of macroparticles injected in the simulation grid), the thermal velocity of electrons in the jet, $v_{\rm jt,th,e} = 0.00014\,c$, and the thermal velocity of electrons in the ambient, $v_{\rm amb,th,e} = 0.05\,c$. The thermal velocity of the ions is smaller than the one of the electrons by a factor $(m_{\rm i}/m_{\rm e})^{1/2}$, where $m_{\rm i}$ is the rest mass of the ion and $m_{\rm e}$ is the rest mass of the electron. This simulation setup is similar to that in the work by \citet{meli23}. 

Similar to Eqs. 26-29 in \citet{hirotani21}, the collision frequency is $\nu_{\rm C} = n_{\rm e} \sigma_{\rm C} v_{\rm e} \sim (\pi n_{\rm e} e^4)/(\gamma^2 m_{\rm e}^2 c^3)$, where $n_{\rm e}$ is the electron concentration, $\sigma_{\rm C}$ is the Coulomb crossection, $v_{\rm e}$ is the jet electron velocity, $e$ and $m_{\rm e}$ denote the charge and the mass of the electrons within the plasma, and $\gamma$ is the electron Lorentz factor. On the other hand, the gyrofrequency is given by $\nu_{\rm B}= (e B)/(2\pi \gamma m_{\rm e} c)$. Thus, the ratio of the collision frequency to the gyrofrequency becomes $\nu_{\rm C}/\nu_{\rm B} \sim (2\pi^2 e^3)/(m_{\rm e} c^2)\cdot \gamma^{-1}\cdot(n_{\rm e}/{\rm cm}^{-3})\cdot(B/1 \rm{G})^{-1} \sim 2.7 \times 10^{-22} \cdot \gamma^{-1}\cdot (n_{\rm e}/{\rm cm}^{-3})\cdot(B/\rm{Gauss})^{-1}$. Taking $\gamma = 15$, $n_{\rm e} \sim 10^4$ cm$^{-3}$ (see the discussions in Section~\ref{conclusions}), and $B \sim 1$ mG, we obtain $\nu_{\rm C}/\nu_{\rm B} \sim 10^{-16}$. We conclude that it is reasonable to neglect Coulomb collisions between particles, and that the plasma in an AGN-like jet is to be regarded as collisionless with anomalous resistivity, and correspondingly it can be treated with PIC method.

Here, we use two types of plasma composition for the jet and the ambient medium: (i) an electron-positron (e$^\pm$) plasma and (ii) an electron-ion (e$^{-}$- i$^{+}$) plasma (with the mass ratio $m_{\rm i}/m_{\rm e} = 4$; note that we have chosen this low mass ratio for ions and electrons to ensure numerical stability for compositions different than the electron-positron plasma). In PIC simulations of relativistic plasma, there is a common practice to use a reduced mass ratio, e.g., $m_{\rm i}/m_{\rm e} = 16$, to
enhance computational efficiency and numerical stability \cite[e.g.,][]{chand24}. To avoid the abnormal growth of MI, we have chosen $m_{\rm i}/m_{\rm e} = 4$, where the ion can be regarded as a heavier positron (see \citet{meli23}). Using such a reduced mass for the ions results in small differences in jet particle acceleration and in the slopes of radiation spectra at low frequencies (see Table\ref{slopes.tab}), in the two cases of plasma compositions.

PIC simulations of magnetized plasma jets with a helical magnetic field have been initially performed by \citet{nishikawa16b}, where the equations for the helical field have been adapted from the work by \citet{mizuno15}. In Cartesian coordinates, the helical magnetic field has the components:
\begin{eqnarray}
B_{x} = \frac{B_{0}}{[1 + (r/a)^2]}, \, B_{\phi} = \frac{(r/a)B_{0}}{[1 + (r/a)^2]}.
\label{B1}
\end{eqnarray}

In the current work, we include only the toroidal component, $B_{\phi}$, of the helical field structure in Eq. (\ref{B1}) (the poloidal component is not included), as in the work by \citet{meli23}. The components of the toroidal magnetic field that is created by a current $+J_{x}(y, z)$ in the positive $x$-direction are:
\begin{equation}
\begin{aligned}
B_{y}(y, z) & =  \frac{((z-z_{\rm jc})/a)B_{0}}{[1 + (r/a)^2]}, \\
B_{z}(y, z) & = -\frac{((y-y_{\rm jc})/a)B_{0}}{[1 + (r/a)^2]}.
\label{B2}
\end{aligned}
\end{equation}
In Eqs.~(\ref{B1}-\ref{B2}), $B_0$ is the amplitude of the initial magnetic field, $a$ is the characteristic length-scale of the toroidal magnetic field, $(y_{\rm jc},\, z_{\rm jc})$ represent the coordinates of the jet center, and $r = [(y-y_{\rm jc})^2+(z-z_{\rm jc})^2]^{1/2}$. The characteristic radius is here set to $a =r_{\rm jt}/4 = 50\Delta$. For more details on the setup of the PIC plasma simulations, see \citet{meli23}. Similar to \citet{meli23}, we have chosen a top-hat density profile for the jet particles to investigate kKHI and MI. Even though this geometry might not be in accordance with real jets, the following step in our future work is to use a jet with a Gaussian density profile to reduce the growth of kKHI and MI. In this case, the kinetic kink instability might grow as in \citet{cerutti23}, which is described in \citet{meli23}. Furthermore, comparisons of such kinetic simulation results with those from MHD simulations could also be beneficial, as kink or any instability able to trigger magnetic turbulence can drive fast reconnection, as predicted in the Lazarian-Vishniac theory \cite[e.g.,][]{vicentin24}, and observed in 3D MHD/PIC simulations of relativistic jets \cite[e.g.,][]{singh16,davelaar20,medina21,medina23}.


Next, we add to the calculations of the PIC plasma code -- with which we obtain physical plasma data as in \citet{meli23} -- new subroutines for calculating the spectra of radiation emitted by the electrons in a jet containing an initial toroidal magnetic field, where we follow the algorithm developed by \citet{hededal05,nishikawa11} of calculating the retarded potentials \citep{jackson99,ryb-light79,hededal05}.

Let us consider a particle located at position ${\rm r}_{0}(t)$, at time $t$.  At the same time, we observe the electric field from the particle at position {\bf r}. However, because of the finite velocity of light, we observe the particle at an earlier position ${\rm r}_{0}(t^{'})$ where it was at the retarded time $t^{'} = t -\delta t^{'} =t -{\bf R} (t^{'} )/c$. Here ${\bf R} (t^{'}) = |{\bf r}- {\bf r}_{0} (t^{'})|$ is the distance from the charge (at the retarded time $t^{'}$) to the observer (at time $t$). After some simplified calculations, the total energy $W$ radiated per unit solid angle per unit frequency from a charged particle moving with instantaneous velocity $\bf{\beta}$ and acceleration $\dot{\bf{\beta}}$ can be expressed as: 

\begin{equation}
  \frac{d^{2}W}{d\Omega d\omega} = \frac{\mu_{0}cq^{2}}{16\pi^{3}} 
            \bigg\vert \int^{\infty}_{-\infty}\frac{{\bf n}\times[({\bf n}-\mathbf{\beta})\times\dot{{\bf \beta}}]}{(1-{\bf \beta}\cdot{\bf n})^{2}}e^{\frac{i\omega(t^{'}-{\bf n}\cdot{\bf r}_{0}(t^{'})}{c}}dt^{'}\bigg\vert^{2}.
 \label{spect}           
\end{equation}
Here, ${\bf n}\equiv {\bf R}(t^{'})/|{\bf R}(t^{'})|$ is a unit vector that points from the retarded position of the particle towards the observer and $q$ is the unit charge. The derivation of the spectral distribution of synchrotron radiation with the full angular dependency can be found in \citet{hededal05}, Appendix C. 

In the calculations presented in the next section, we parameterized the initial toroidal magnetic field (Eq.~\ref{B2}) through its initial amplitude $B_0$, using two values: one for a moderate magnetic field strength, $B_0=0.5$, as in the work by \citet{meli23}, and the other one for a weaker magnetic field strength, $B_0=0.1$. 
These two values of the amplitude of the initial toroidal field, $B_0=0.5$ and $B_0=0.1$, correspond to plasma magnetization, in the observer frame, $\sigma = B_{\rm 0}^{2}/(n_{\rm e}m_{\rm e}\Gamma c^{2})$ of $\sim 6.92\times 10^{-4}$ and $\sim 1.73\times 10^{-2}$, respectively, in the case of a $\Gamma = 15$ jet. For a  $\Gamma = 100$ jet, the corresponding values of magnetizations are $\sim 2.6\times 10^{-1}$ and $\sim 1.3\times 10^{-4}$. (The magnetization parameters are calculated in simulation units). The magnetization parameter of the plasma varies along the jet, increasing locally in regions where the magnetic field becomes stronger due to kinetic instabilities.

For each of them, we calculate spectra from electrons accelerated in a relativistic plasma jet that propagates with a bulk Lorentz factor of $\Gamma=15$ (which is typical for a jet in AGN) and of $\Gamma=100$ (which is typical for a GRB jet). The quantity $\Gamma$ denotes the product $\beta \Gamma_0$, with $\beta = v/c$ and $\Gamma_0 = (1 - \beta^2 )^{-1/2}$, the jet Lorentz factor, where $v$ is the jet velocity and $c$ is the speed of light. There is a common practice to denote $\Gamma$ as the (bulk) Lorentz factor of the jet instead of $\Gamma_0$. In addition, for each of these setups of the jet Lorentz factor, we consider the two species for the jet plasma: e$^\pm$ and e$^{-}$- i$^{+}$ (with $m_{\rm i}/m_{\rm e} = 4$), respectively. We note that for each specific setup, the jet plasma and the ambient plasma have the same kind of composition, either e$^\pm$ or e$^{-}$- i$^{+}$. (Different plasma compositions for the jet and for the ambient, in one particular setup, should be used in further work.) Therefore, we obtain eight sets of radiation spectra. Furthermore, each set of radiation spectra is composed of two parts: one for a head-on ($\theta=0^\circ$) emission of radiation and the other one for a 5$^\circ$-off axis emission of radiation (here, $\theta$ is the angle between the axis of the jet and the observer's line of sight).

\section{Results and interpretation}\label{results}

In this section, we study the spectra from a {\it top-hat} nonthermal distribution of particles moving in the electromagnetic fields modified by kinetic instabilities in the presence of an initial toroidal magnetic field. Before determining the radiation spectra (Subsect.~\ref{sec:spectra}), we present the global system evolution and electron acceleration (Subsect.~\ref{evolution}) and perform an analysis of turbulence (Subsect.~\ref{fft}).

\subsection{Global system evolution and electron acceleration}
\label{evolution}

\begin{figure}
\begin{center}
\hspace*{0.0cm} {\bf e$^{\pm}$ jet with $\mathbf{\Gamma=15}$} \hspace*{0.0cm} (a) \hspace*{0.0cm} {\bf e$^{-}$- i$^{+}$ jet with $\mathbf{\Gamma=15}$} \hspace*{0.0cm} (b) 

\includegraphics[scale=0.3,angle=0]{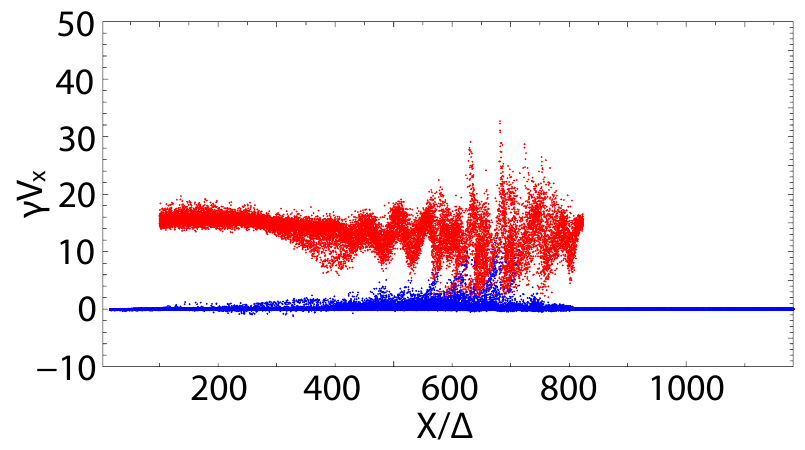}
\includegraphics[scale=0.3,angle=0]{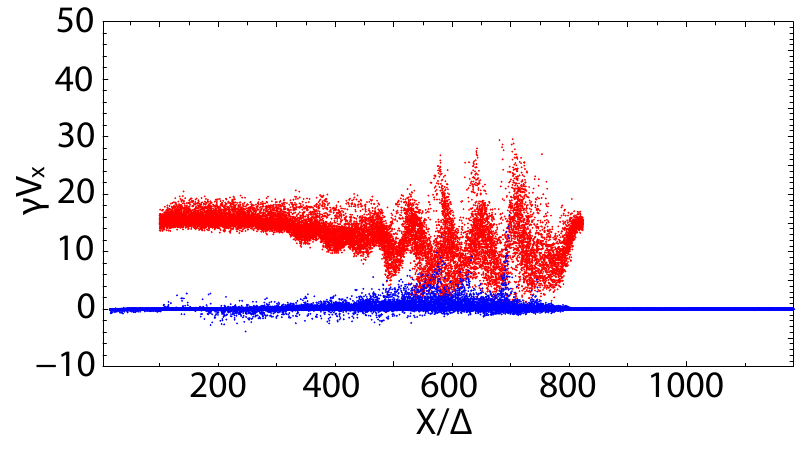}

\hspace*{0.0cm} {\bf e$^{\pm}$ jet with $\mathbf{\Gamma=100}$} \hspace*{0.0cm} (c) \hspace*{0.0cm} {\bf e$^{-}$- i$^{+}$ jet with $\mathbf{\Gamma=100}$} \hspace*{0.0cm} (d) 

\includegraphics[scale=0.3,angle=0]{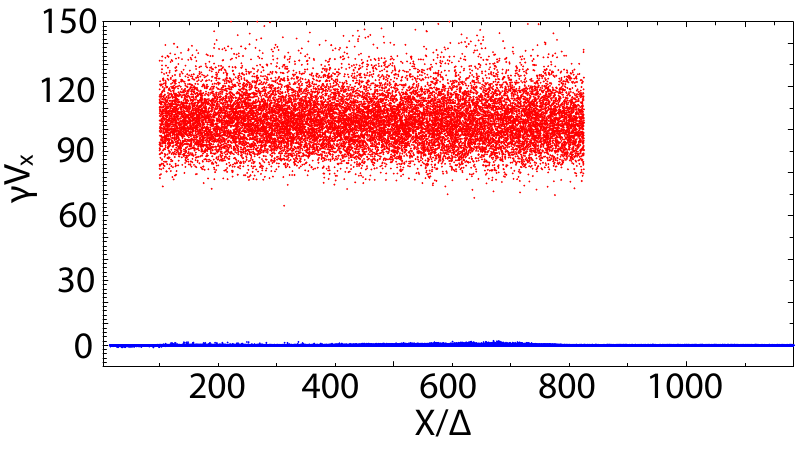}
\includegraphics[scale=0.3,angle=0]{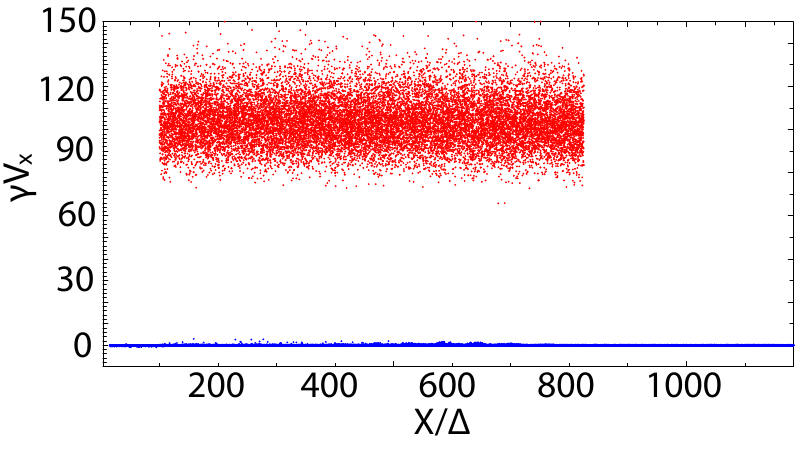}
\end{center}
\vspace{-0.5cm}
\caption{$x$ - $\gamma v_{x}$ distribution of jet (red) and  ambient (blue) electrons for an e$^{\pm}$ jet (left panels) and for an e$^{-}$- i$^{+}$ jet (right panels) with $B_{0}=0.5$, at $t = 725\omega^{-1}_{\rm pe}$. 
Upper panels correspond to jets with a jet Lorentz factor of $\Gamma=15$, while lower panels are for jets with $\Gamma=100$.
} 
\label{LorM}
\end{figure}

\begin{figure}
\begin{center}
\hspace*{0.0cm} {\bf e$^{\pm}$ jet with $\mathbf{\Gamma=15}$} \hspace*{0.0cm} (a) \hspace*{0.0cm} {\bf e$^{-}$- i$^{+}$ jet with $\mathbf{\Gamma=15}$} \hspace*{0.0cm} (b) 

\includegraphics[scale=0.3,angle=0]{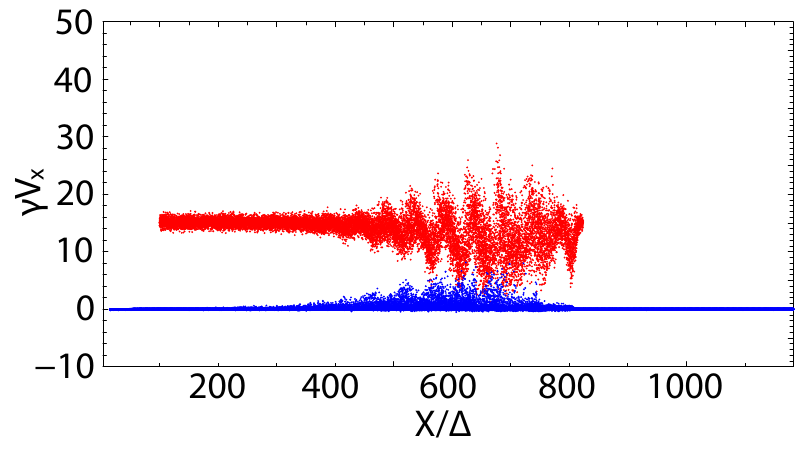}
\includegraphics[scale=0.3,angle=0]{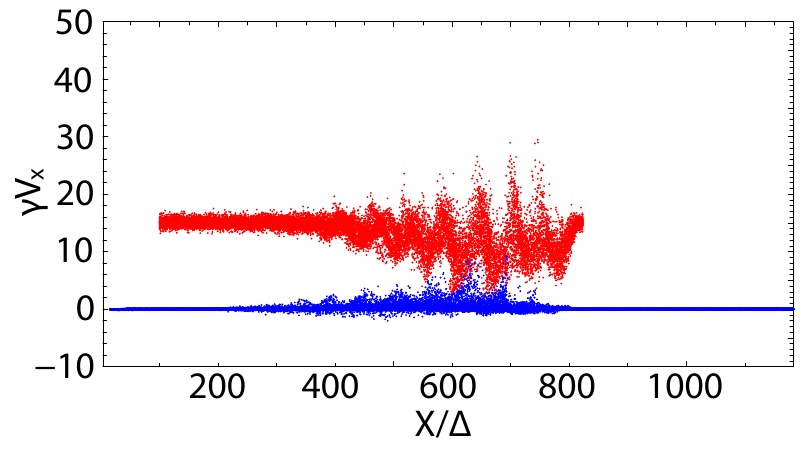}

\hspace*{0.0cm} {\bf e$^{\pm}$ jet with $\mathbf{\Gamma=100}$} \hspace*{0.0cm} (c) \hspace*{0.0cm} {\bf e$^{-}$- i$^{+}$ jet with $\mathbf{\Gamma=100}$} \hspace*{0.0cm} (d) 

\includegraphics[scale=0.3,angle=0]{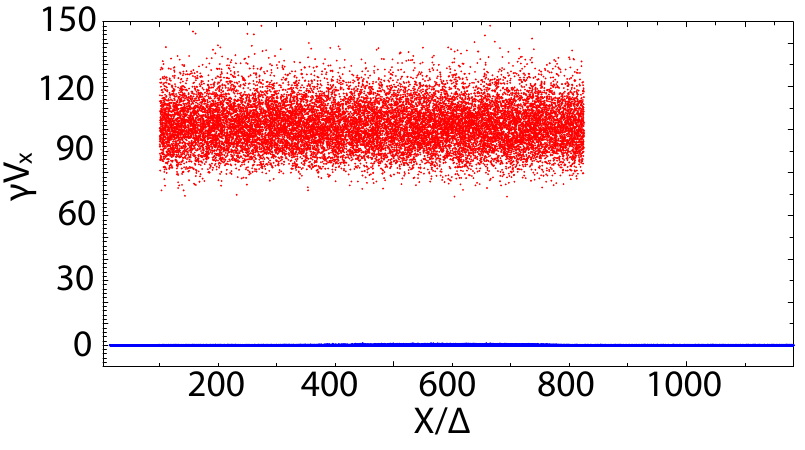}
\includegraphics[scale=0.3,angle=0]{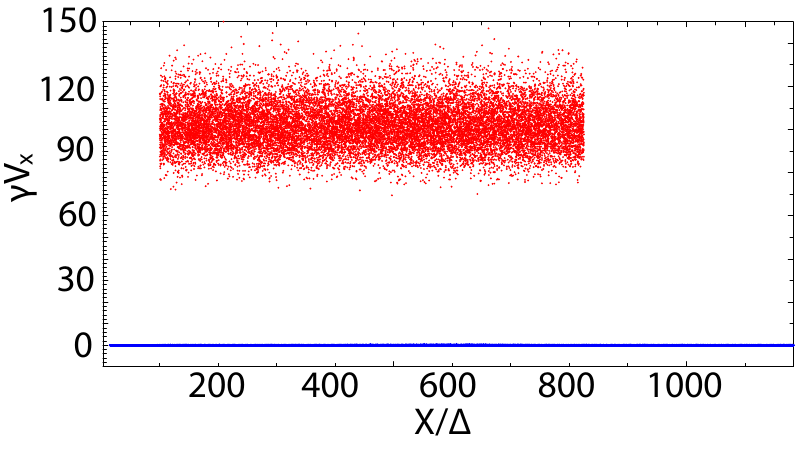}
\end{center}
\vspace{-0.5cm}
\caption{$x$ - $\gamma v_{x}$ distribution of jet (red) and  ambient (blue) electrons for an e$^{\pm}$ jet (left panels) and for an e$^{-}$- i$^{+}$ jet (right panels) with $B_{0}=0.1$, at $t = 725\omega^{-1}_{\rm pe}$. Upper panels correspond to jets with a bulk Lorentz factor of $\Gamma=15$, while lower panels are for jets with $\Gamma=100$.} 
\label{LorW}
\end{figure}

\begin{figure*}
\begin{minipage}{14cm}
\begin{center}
\hspace*{2.0cm} {\bf e$^{\pm}$ jet with $\mathbf{\Gamma=15}$} \hspace*{1.5cm} (a) \hspace*{2.4cm} {\bf e$^{-}$- i$^{+}$ jet with $\mathbf{\Gamma=15}$} \hspace*{1.5cm} (b) 
\includegraphics[scale=0.39,angle=0]{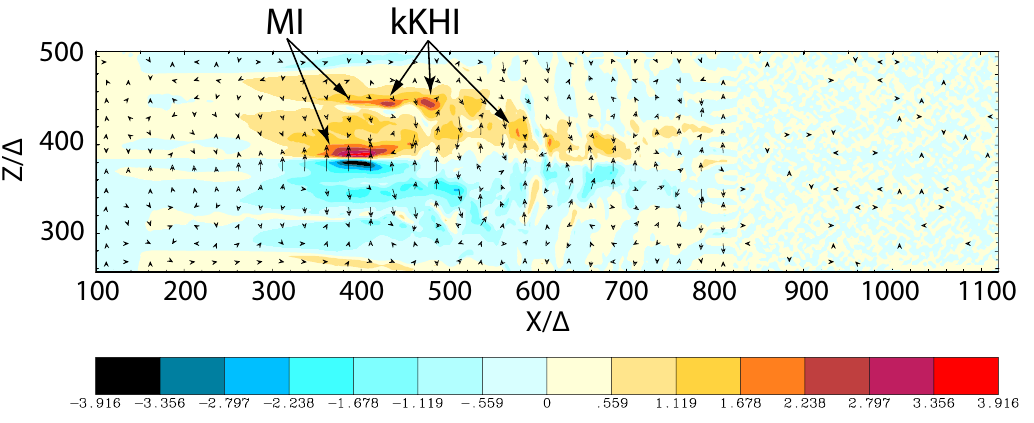}
\includegraphics[scale=0.39,angle=0]{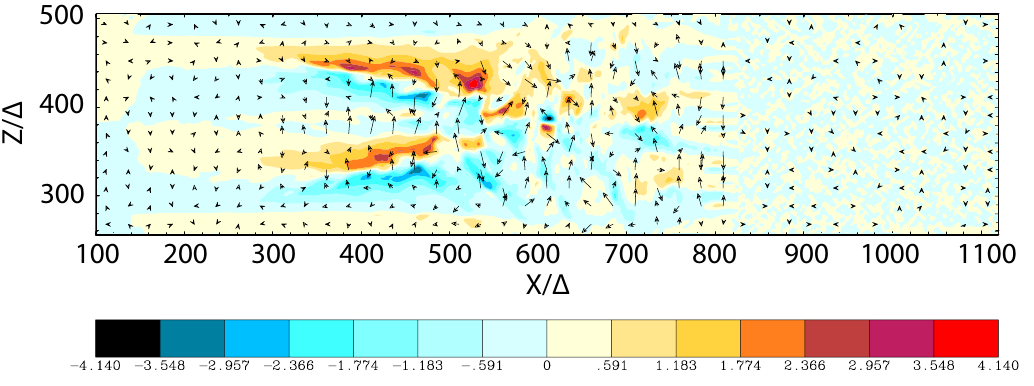}

\hspace*{2.0cm} {\bf e$^{\pm}$ jet with $\mathbf{\Gamma=100}$} \hspace*{1.5cm} (c) \hspace*{2.4cm} {\bf e$^{-}$- i$^{+}$ jet with $\mathbf{\Gamma=100}$} \hspace*{1.5cm} (d) 
\includegraphics[scale=0.39,angle=0]{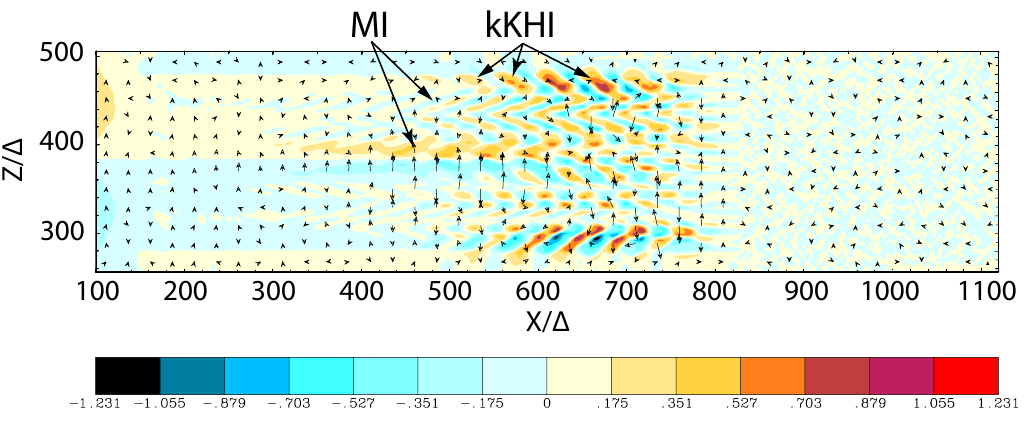}
\includegraphics[scale=0.39,angle=0]{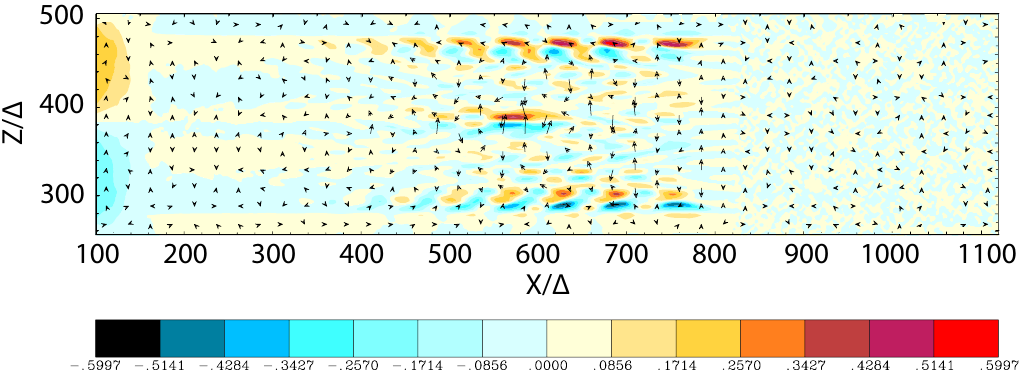}
\end{center}
\end{minipage}
\begin{minipage}{3.5cm}
\vspace{-0.5cm}
\caption{Distribution of the magnetic field in the $x-z$ plane of the simulation box, given for the case with $B_{0} = 0.5$, at $t = 725\,\omega^{-1}_{\rm pe}$.
The left panels (a, c) are for an e$^{\pm}$ jet, the right panels (b, d) are for an e$^{-}$ - i$^{+}$ jet.
The upper panels (a, b) correspond to jets with a bulk Lorentz factor of $\Gamma=15$, the lower panels (c, d) are for jets with $\Gamma=100$.
The out-of-plain $B_{y}$ component is shown with color maps, the in-plain field combined of $B_{x}$ and $B_{z}$ components is shown with arrows.
The maximum and minimum of $B_{\rm xz}$ are (a): $\pm 3.916$, (b): $\pm 4.140$, (c): $\pm 1.231$, and (d): $\pm 0.5997$.}
\label{BFM}
\end{minipage}
\end{figure*}

\begin{figure*}
\begin{minipage}{14cm}
\begin{center}
\hspace*{2.0cm} {\bf e$^{\pm}$ jet with $\mathbf{\Gamma=15}$} \hspace*{1.5cm} (a) \hspace*{2.4cm} {\bf e$^{-}$- i$^{+}$ jet with $\mathbf{\Gamma=15}$} \hspace*{1.5cm} (b) 
\includegraphics[scale=0.39,angle=0]{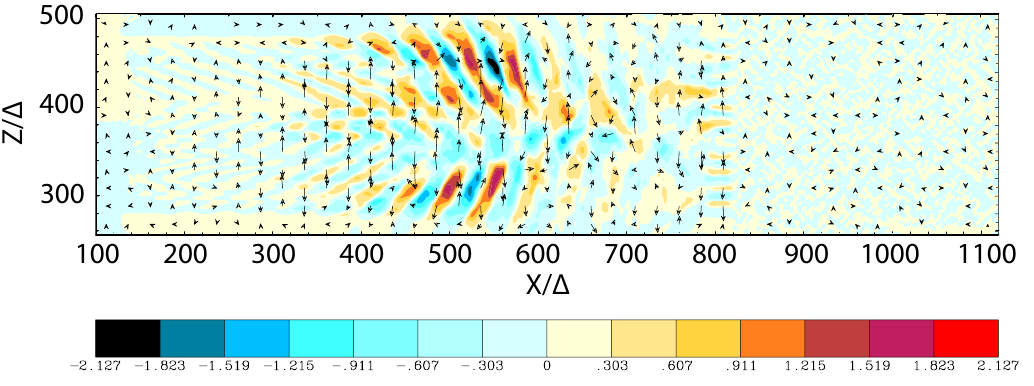}
\includegraphics[scale=0.39,angle=0]{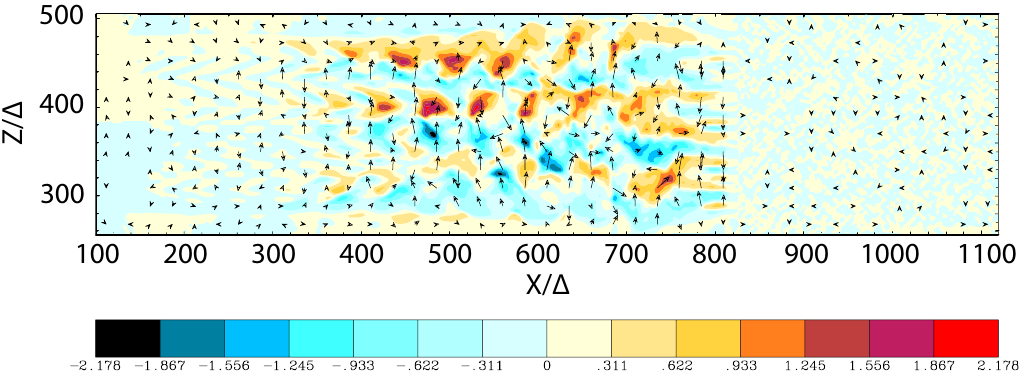}

\hspace*{2.0cm} {\bf e$^{\pm}$ jet with $\mathbf{\Gamma=100}$} \hspace*{1.5cm} (c) \hspace*{2.4cm} {\bf e$^{-}$- i$^{+}$ jet with $\mathbf{\Gamma=100}$} \hspace*{1.5cm} (d) 
\includegraphics[scale=0.39,angle=0]{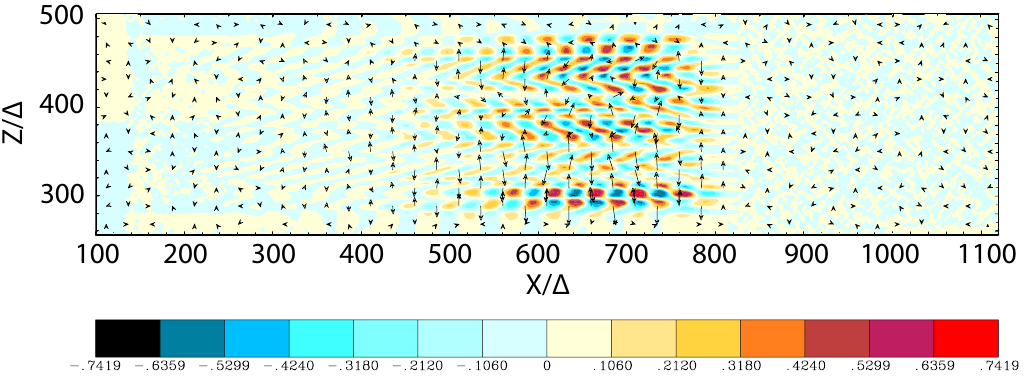}
\includegraphics[scale=0.39,angle=0]{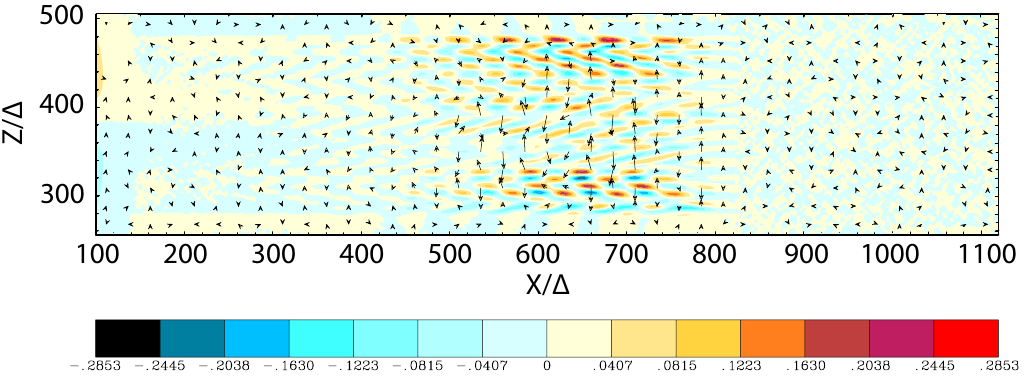}
\end{center}
\end{minipage}
\begin{minipage}{3.5cm}
\vspace{-0.5cm}
\caption{Distribution of the magnetic field in the $x-z$ plane of the simulation box, given for the case with $B_{0} = 0.1$, at $t = 725\,\omega^{-1}_{\rm pe}$.
The left panels (a, c) are for an e$^{\pm}$ jet, the right panels (b, d) are for an e$^{-}$ - i$^{+}$ jet.
The upper panels (a, b) correspond to jets with a bulk Lorentz factor of $\Gamma=15$, the lower panels (c, d) are for jets with $\Gamma=100$.
The out-of-plain $B_{y}$ component is shown with color maps, the in-plain field combined of $B_{x}$ and $B_{z}$ components is shown with arrows.
The maximum and minimum of $B_{\rm xz}$ are (a): $\pm 2.127$, (b): $\pm 2.178$, (c): $\pm 0.7419$, and (d): $\pm 0.2853$.
}
\label{BFW}
\end{minipage}
\end{figure*}

\begin{figure}
\begin{center}

\includegraphics[scale=0.54,angle=0]{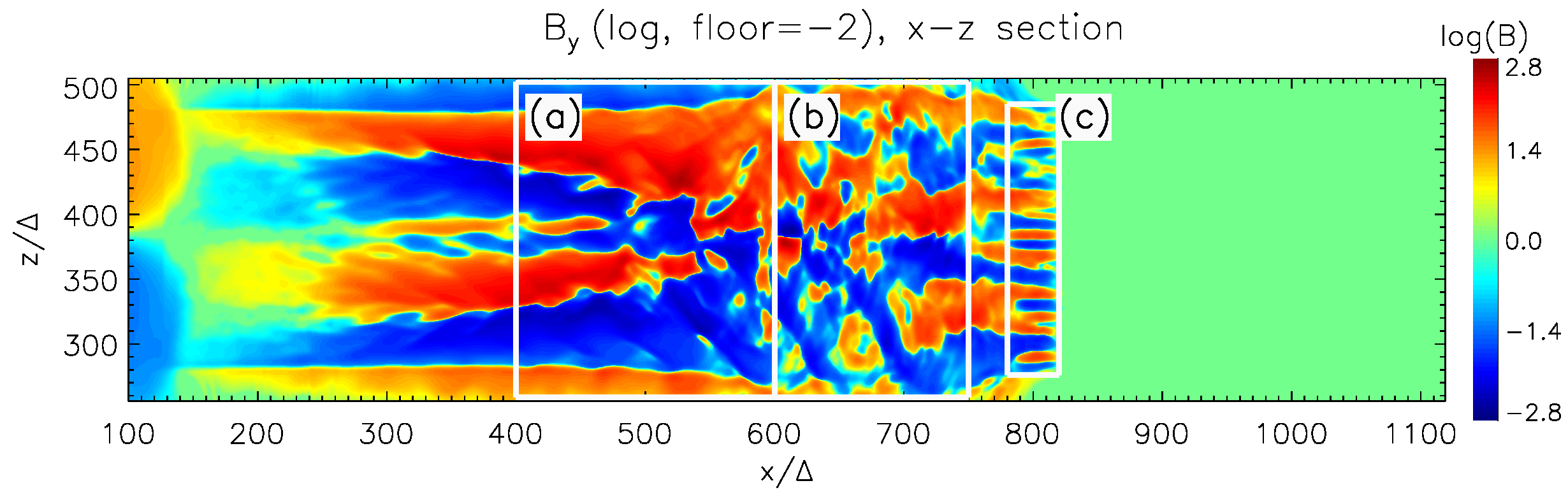}

\end{center}
\caption{The same plot as Fig.~\ref{BFM}(b), but with use of the sign-preserving logarithmic color scaling for $B_{y}$. We plot $\mathrm{sgn}(B_{y}) \cdot \{2+\log[\max(|B_{y}|/B_{0},10^{-2})]\}$. The level of "0" on the color scale hence corresponds to $|B|/B_{0} \le 10^{-2}$.
The regions (a), (b) and (c) marked with white rectangles were used for FFT plots presented in Fig.~\ref{B_fft}.}
\label{Bylog}
\end{figure}

\begin{figure*}
\begin{center}
\hspace*{0cm}  (a) \hspace*{5.0cm}  (b) \hspace*{5.0cm} (c) 

\includegraphics[scale=0.42,angle=0]{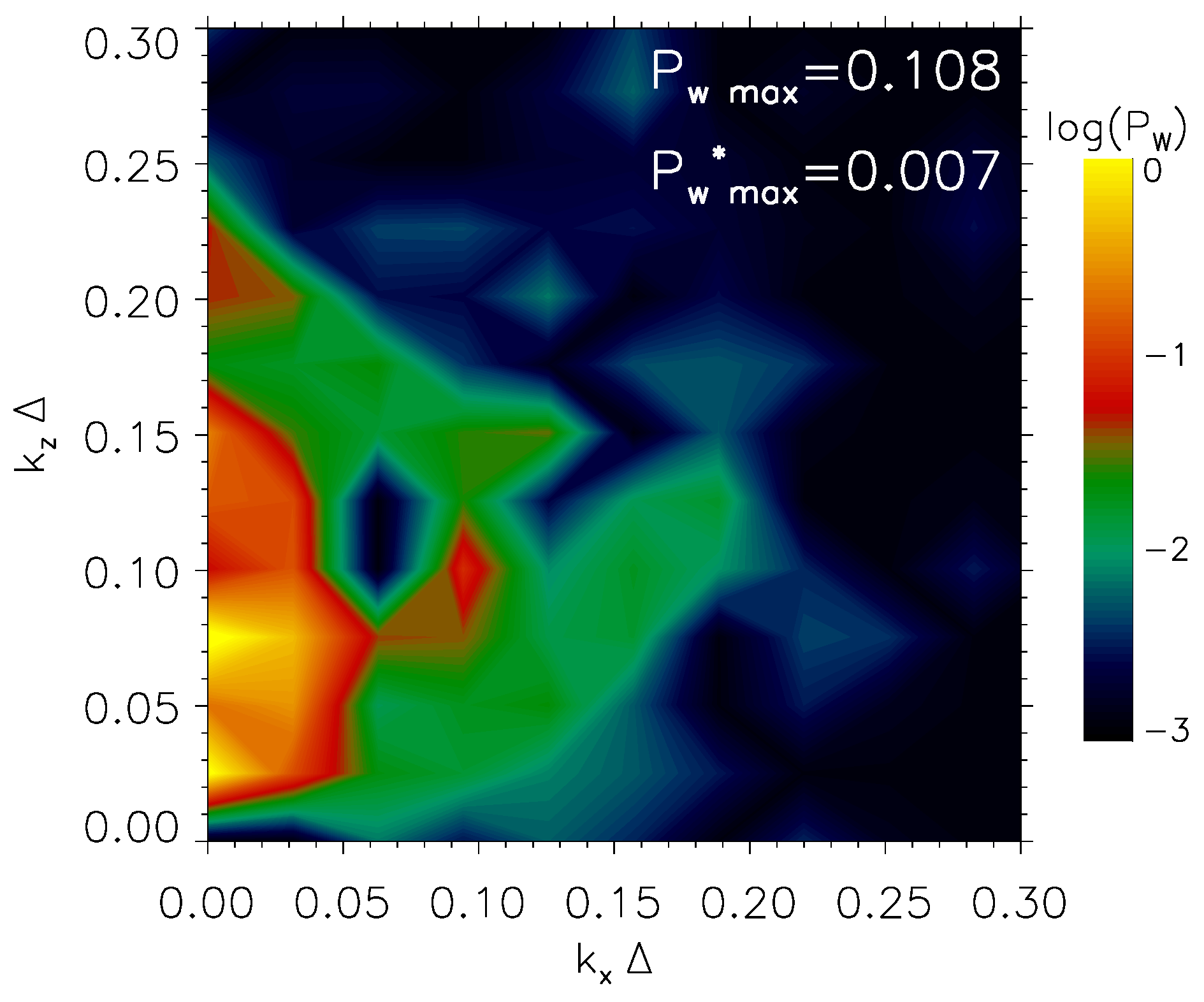}
~~~~~
\includegraphics[scale=0.42,angle=0]{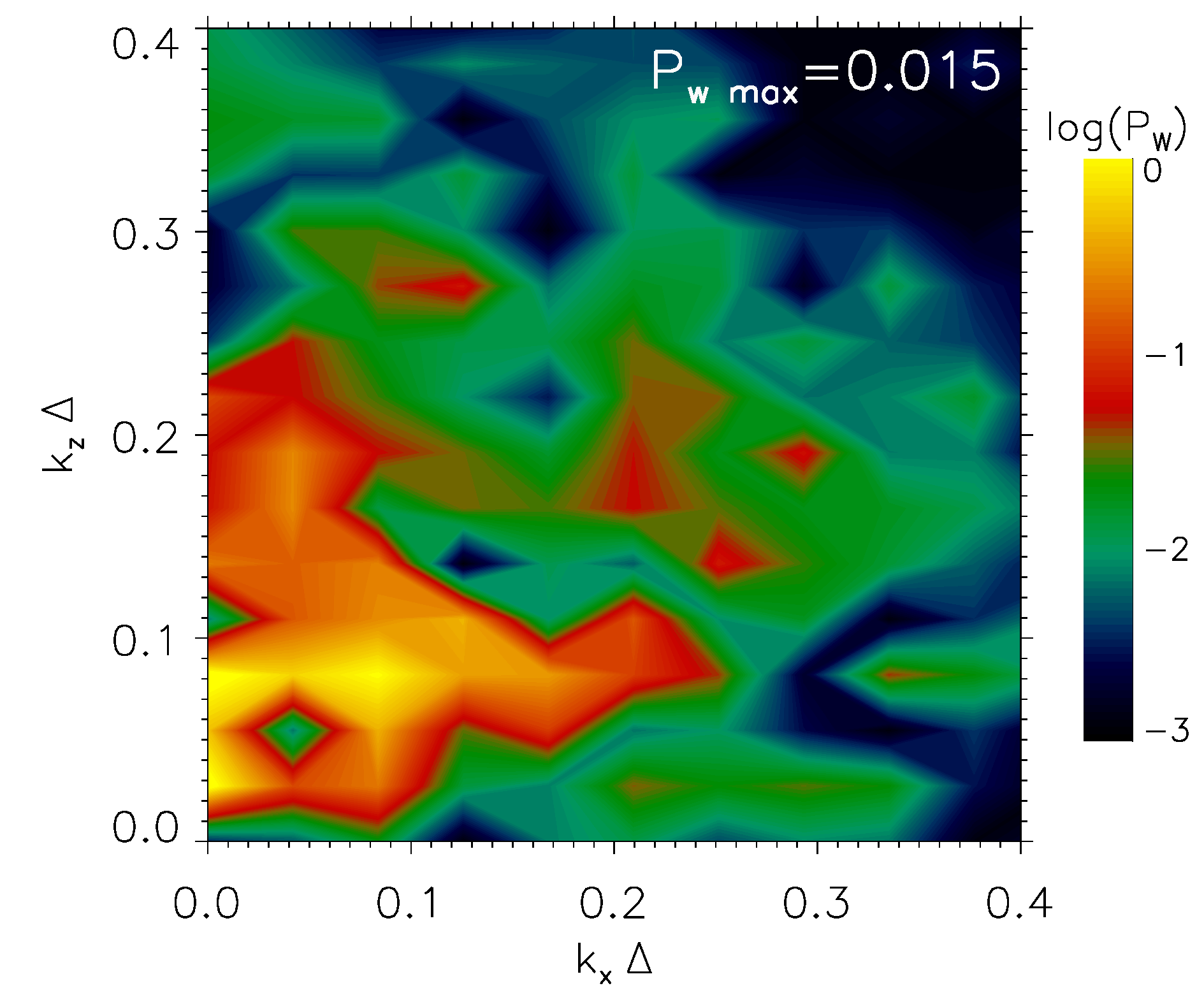}
~~~~~
\includegraphics[scale=0.42,angle=0]{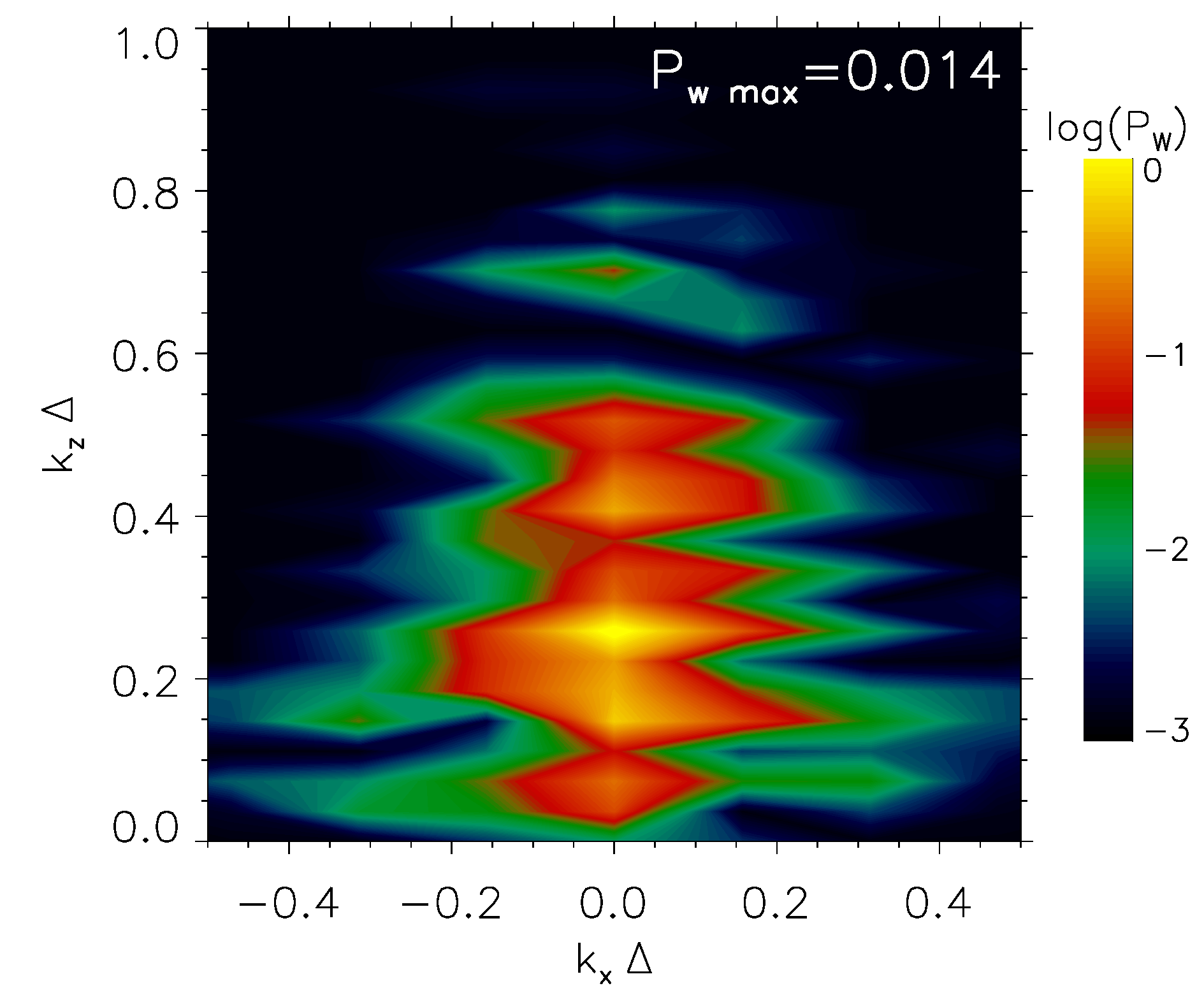}

\end{center}
\caption{FFT-images of the magnetic field distribution ($B_{y}$ component) shown in Fig.~\ref{BFM}(b) for the e$^{-}$- i$^{+}$ jet with $\Gamma=15$ and $B_{0}=0.5$. The panels represent three regions: (a) $x\in[400-600]\,\Delta$, (b) $x\in[600-750]\,\Delta$ and (c) $x\in[780-820]\,\Delta$, which are also marked with white rectangles in Fig.~\ref{Bylog}.
The color scale is logarithmic and normalized to the maximum wave power, so that numbers at the color bar mean $\rm{log}(P_{\rm{w}}/P_{\rm{w~max}})$. Corresponding values of $P_{\rm{w~max}}$ are given in right upper corners. $P^{~*}_{\rm{w~max}} = 0.007$ in panel (a) corresponds to the local maximum at $(k_{x},\,k_{z})\,\Delta \approx (0.095,\,0.1)$. These FFT-images are used for estimating the characteristic length-scale of the turbulence in the jet plasma.}
\label{B_fft}
\end{figure*}

The $x$ - $\gamma v_{x}$ distribution of the jet (red) and ambient (blue) electrons are shown for $B_{0}=0.5$ (in Fig.~\ref{LorM}) and for $B_{0}=0.1$ (in Fig.~\ref{LorW}) for an e$^{\pm}$ and for an e$^{-}$- i$^{+}$ jet (right panels) with initial jet Lorentz factors of $\Gamma=15$ and $\Gamma=100$. Although we do not calculate the spectra produced by the ambient electrons, we include here their phase-space ($x$ - $\gamma v_{x}$) representation just for comparison with the phase-space distribution of the electrons accelerated in the jet.

For a plasma jet with initial $\Gamma = 15$, Figs.~\ref{LorM}(a,b) and ~\ref{LorW}(a,b) show that the jet electrons (in red) are accelerated in bunches by growing instabilities; hence their momenta along the $x$-direction reach high values, where $\gamma {\rm v}_{x}$ doubles its value from $\gamma v_{x} \sim 15$ at injection, up to $\gamma v_{x} \sim 35$. In contrast, for a plasma jet with initial $\Gamma = 100$, the initial momentum of the jet electrons increases only by maximally 50$\%$ (Figs.~\ref{LorM}(c,d) and ~\ref{LorW}(c,d)).

In PIC simulations we can distinguish between the jet and the ambient particles, however physically these two populations of particles can mix through interactions. The ambient electrons in or near the jet are accelerated through kinetic instabilities, and they can be identified by the spikes in the $x-\gamma v_x$ distribution, in blue color, in Figs.~\ref{LorM}(a,b) and \ref{LorW}(a,b), whereas the ambient electrons outside the jet are not accelerated.

In Fig.~\ref{LorW}, both the difference in the strength of the amplitude of the applied toroidal magnetic field by a factor of 5 and the ion-to-electron mass ratio of 4 have a weak impact on the $x - \gamma v_x$ distribution of the jet electrons, i.e., on acceleration of jet electrons. Nevertheless, some differences in the slopes of the spectra can be seen in Table~\ref{slopes.tab}. In the case of a jet with $\Gamma=100$, the peak intensity of the radiation increases by one order of magnitude as the strength of the amplitude of the initial toroidal magnetic field increases by a factor of 5.

The role of kinetic instabilities on particle acceleration is thoroughly described in \citet{meli23}. In the linear regime of the plasma, the WI grows first, then kKHI and MI grow until the MI becomes dominant over the kKHI. In the current paper, we show the evolution of the plasma at the nonlinear stage (simulation time $t=725 \omega_{\rm pe}^{-1}$), where the jet particles are accelerated possibly due to reconnection (as observed in \citet{meli23}).The quasi-stationary parallel electric field mentioned here primarily refers to the acceleration occurring in region (b) of Fig.~\ref{Bylog}, following the nonlinear and turbulent phases. After the end of the nonlinear stage, a quasi-stationary parallel electric field accelerates jet electrons further (see \citet{meli23}). We emphasize that multiple mechanisms, including reconnection-driven and Fermi-like processes, may be at play, with their relative importance requiring further investigation.

In Fig.~\ref{BFM}(a), around $x \sim 400 \Delta$, we observe that the growth of MI\footnote{From the time evolution of the magnetic field provided by the supplemental material in \citet{meli23}, one can see that the MI dominates over the WI.} is stronger, which in the $x-z$ plane can be seen as two pairs of layers, with opposite polarity of the magnetic field (these layers are actually the projection onto the $x-z$ plane of the circularly clockwise - as seen from the head of the jet - $B_{\phi}$ component of the magnetic field generated by the MI). At about $x \sim 500 \Delta$, the magnetic field generated by MI is modulated by the kKHI, which later on dominates over the MI (as its growth time is longer than that of MI). Then, the nonlinear regime of plasma instabilities is reached, and the magnetic fields starts to dissipate. This is the stage of the plasma magnetic field when we begin to select jet electrons to calculate spectra. In Fig.~\ref{BFM}(b), the instabilities grow in a similar way as in the case in Fig.~\ref{BFM}(a), except that the kKHI is more visible.

For a weaker amplitude of the initial toroidal magnetic field (i) in Figs.~\ref{BFW}(a,c,d) the dominant growing mode of kinetic instabilities is the WI, depicted as oblate stripes, which is different from the cases in Figs.~\ref{BFM}(a,c,d), where the modes of MI and kKHI are dominant and (ii) in Fig.~\ref{BFW}(b), the kKHI and MI are more visible than in Figs.~\ref{BFW}(a,c,d).

In Fig.~\ref{Bylog}, which is the same as plot Fig.~\ref{BFM}(b), but with the use of the sign-preserving logarithmic color scaling for $B_y$, we separate different regimes based on the structures of the observed wave modes; that is, region (a) represents the nonlinear stage, region (b) represents the turbulent plasma stage, and region (c) is behind the jet head containing wave vectors that are perpendicular to the jet direction (see Fig.~\ref{B_fft}c). To the left of the region (a), the linear growth of instabilities takes place. The striping structures in region (c) are formed by a double layer plasma, and they are associated with an ambipolar electrostatic field \citep{ardaneh16}. Based on Fig.~\ref{Bylog}, due to the saturation of the nonlinear regime, a turbulent magnetic field is generated, the size of which is 1/4-1/3 of the jet radius.

\subsection{Analysis of turbulence}
\label{fft}

We use the Fast Fourier Transform (FFT) to convert the amplitude of the $B_{y}$ wave components of the magnetic field shown in Fig.~\ref{BFM}(b) from the spatial domain to the wave vector domain, in order to estimate the characteristic length-scale of the turbulence. Figure~\ref{Bylog} represents the same distribution of $B_{y}$ as in Fig.~\ref{BFM}(b), but plotted in logarithmic color scale, in order to resolve weaker field amplitudes together with stronger ones. Such an approach helps us to make the proper selection of the space regions for FFT analysis. These regions are marked in Fig.~\ref{Bylog} with white rectangles: (a) for $x \in [400 - 600]\,\Delta$, (b) for $x \in [600 - 750]\,\Delta$ and (c) for $x \in [780 - 820]\,\Delta$. The selection has been made in a way to avoid the mixture of different waves in the same region, to make the FFT plots cleaner and easier for interpretation.

In Figure~\ref{B_fft} we show the power spectra of the waves excited by kinetic plasma instabilities, computed for the regions (a), (b) and (c) marked in Fig.~\ref{Bylog}. These spectra are plotted in logarithmic color scale, in the ($k_{x}$, $k_{z}$) wave vector space, where the wave vector is defined as $|\vec{k}| \equiv 2\pi/\lambda$.

\textbf{\textit{Region~(a)}}.
The strongest modes observed in Fig.~\ref{B_fft}(a) are perpendicular to the jet axis ($k_{x}\,\Delta = 0$) and have two maxima at $k_{z}\,\Delta = 0.025 \pm 0.007$ and $k_{z}\,\Delta = 0.075 \pm 0.007$. The corresponding wavelengths $\lambda = (250 \pm 70)\,\Delta$ and $\lambda = (84 \pm 8)\,\Delta$ are related rather to the global radial structure of the jet than to the turbulence. At the same time, there is a weaker oblique mode at $(k_{x},\,k_{z})\,\Delta = (0.095 \pm 0.009,\,0.100 \pm 0.007)$, with corresponding wavelength $\lambda_{\rm{obl}} = 2 \pi/\sqrt{k_{x}^2 + k_{z}^2} = (45 \pm 4)\,\Delta$ and relative wave power $P_{\rm w} \approx 0.007$. The later is associated with turbulent structure in this region with obliquity $\theta \approx 45^\circ$.

\textbf{\textit{Region~(b)}}.
This region is evidently most turbulent among all selected (see Fig.~\ref{Bylog}). Correspondingly, a mixture of various wave modes in this region is observed in the FFT plot in Fig.~\ref{B_fft}(b). The dominating modes are two perpendicular with $k_{z}\,\Delta = 0.028 \pm 0.007$ and $k_{z}\,\Delta = 0.080 \pm 0.007$, as well as one oblique with $(k_{x},\,k_{z})\,\Delta = (0.085 \pm 0.012,\,0.082 \pm 0.007)$. The corresponding wavelengths are $\lambda = (250 \pm 70)\,\Delta$, $\lambda = (79 \pm 7)\,\Delta$ and $\lambda = (53 \pm 6)\,\Delta$, in the same sequence. The first one, like in the case of region (a), reflects the global system symmetry and it is not related to the turbulence. Meanwhile, two other with $\lambda_{\perp} \approx 79\,\Delta$ and $\lambda_{\rm{obl}} \approx 53\,\Delta$ ($\theta \approx 45^\circ$) are obviously related to the strong magnetic turbulence observed in this region and have comparable wave power $P_{\rm w} \approx 0.015$.

\textbf{\textit{Region~(c)}}
is relatively thin and corresponds to the front of the jet. 
Because of its small thickness, only the perpendicular modes can be distinguished here with FFT.  The striping structures in this region are formed by a double layer plasma, and they are associated with an ambipolar electrostatic field \citep{ardaneh16}. In Fig.~\ref{B_fft}(c) these modes are presented with a set of maxima at $k_{x}\,\Delta = 0$. The strongest one has $k_{z}\,\Delta = 0.26 \pm 0.01$ corresponding to a wavelength of $\lambda_{\perp} = (24 \pm 1)\,\Delta$ with relative wave power $P_{\rm w} \approx 0.014$. Other maxima are several times weaker.

We note that in our simulations, the electron skin depth is $\lambda_{\rm e} = 10\,\Delta$, therefore the wavelength of the strongest wave mode becomes $\lambda_{\perp} = (2.4 \pm 0.1)\,\lambda_{\rm e}$.  For comparison, in the theory of the jitter radiation \cite[e.g.,][]{medvedev99,medvedev00}, the characteristic coherence scale of the generated magnetic field by the relativistic generalization of the two-stream WI in an electron-ion plasma is of the order of the relativistic skin depth, $\sim \lambda_{\rm e}$. Furthermore, the front region where $x\in[600-800]\,\Delta$ might correspond to the jet region from where we trace the electrons for calculating the spectra of jitter-like radiation (see the next subsection).

\subsection{Synthetic spectra calculation}
\label{sec:spectra}

\begin{figure}
	\centering
	\includegraphics[width=0.7\columnwidth]{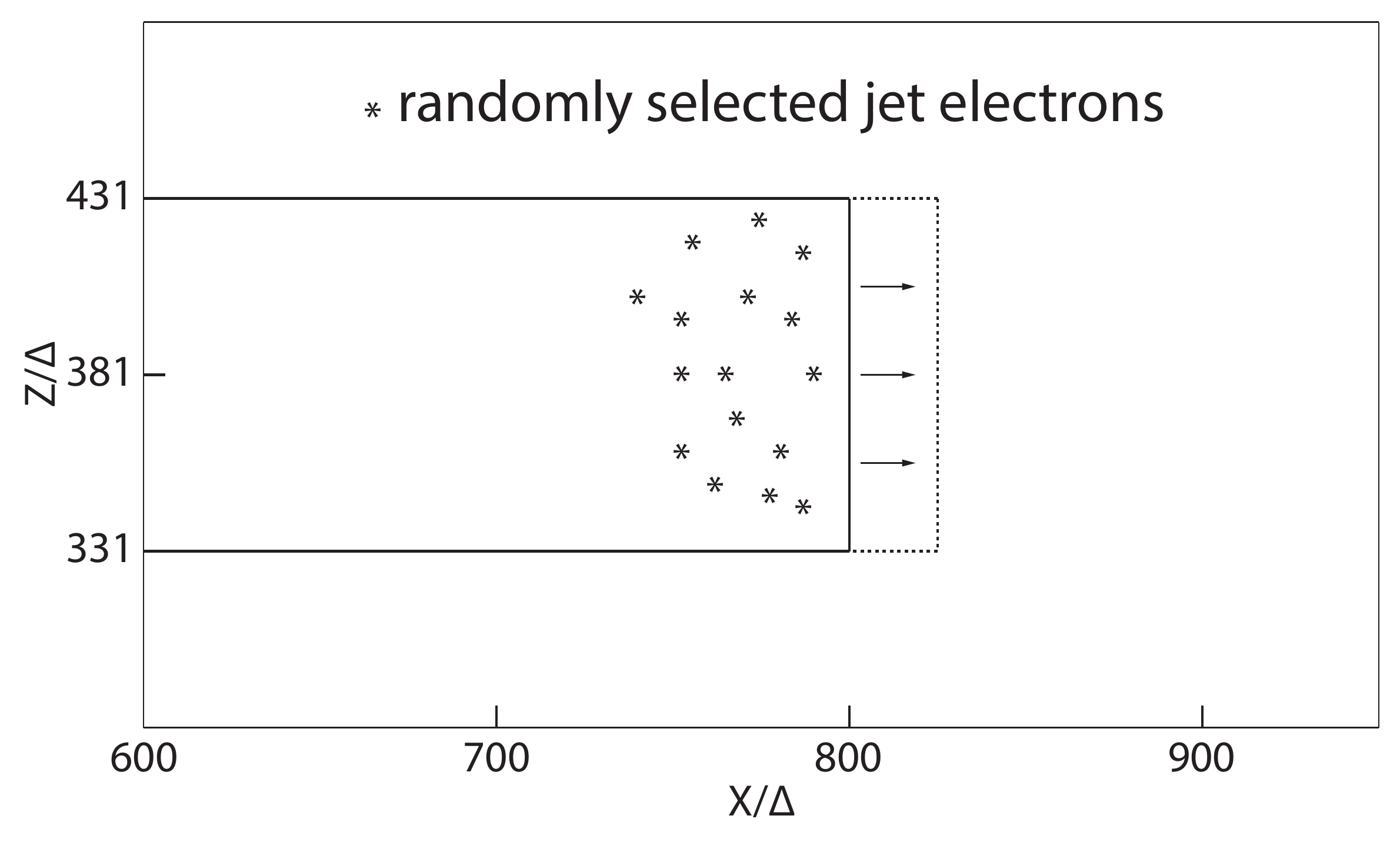}	
	\vspace{-0.5cm}
	\caption{Schematic representation of the frontal jet volume, from where the electrons are selected when performing calculations of the radiation spectra. During the spectra calculations, the jet head moves 25 grid cells, from $x = 800 \Delta$ to $x = 825 \Delta$.  We trace jet electrons that have a Lorentz factor of $\gamma>12$ and $\gamma>80$ for a plasma jet moving with a bulk Lorentz factor of $\Gamma = 15$ and $\Gamma = 100$, respectively.}
	\label{scheme}
\end{figure}

We note that $t = 700\, \omega_{\rm pe}^{-1}$ represents a time in the simulation well into the non-linear regime of the developing instabilities (see \cite{meli23}). By this simulation time, the head of the jet has reached $x = 800 \Delta$.

To obtain a radiation spectrum from PIC simulations at a given time $t$, we have to trace the jet electrons over a time interval around $t$ with a high temporal resolution $\Delta t_{\rm s}$. It is near to impossible to trace all jet electrons and, therefore, we select (or sample) a feasible number of electrons, maximum 8000 (see more details a few paragraphs below). Once the positions, velocities, and accelerations of the jet electrons are sampled, the radiation spectra can be computed using Eq. (\ref{spect}). 

We randomly select jet electrons from a region situated behind the jet head, as indicated by the small stars in Fig.~\ref{scheme}, to calculate the resulting spectra. Throughout this section, we refer to the selected sample of jet electrons as jet electrons when we mention synthetic spectra calculations.  During the spectra calculations, the jet head moves 25 grid cells, from $x = 800 \Delta$ to $x = 825 \Delta$. We note that the jet particles are accelerated by the quasi-stationary parallel electric field generated due to the growing kKHI and MI, as described in \citet{meli23} (see Figs. 10(c,d) in \citet{meli23}). Behind the jet head, some of the jet electrons are additionally accelerated by a quasi-stationary parallel electric field. It may be possible that the acceleration mechanism in the region where turbulence driven by kinetic instabilities dominates and reconnection layers may be present to be similar to the turbulence-induced reconnection acceleration described in the MHD studies  \cite[e.g.,][]{kowal12,delValle16,medina21,medina23,deGouveia24}. However, a more detailed quantitative analysis, including the computation of the acceleration rate and the properties of the reconnection layers driven by the instabilities and turbulence, is subject of future work to enable more precise comparisons with the MHD simulations.

\begin{figure*}
\begin{center}
\hspace*{2.0cm} {\bf e$^{\pm}$ jet with $\mathbf{\Gamma=15}$} \hspace*{1.7cm} (a) \hspace*{3.2cm} {\bf e$^{-}$- i$^{+}$ jet with $\mathbf{\Gamma=15}$} \hspace*{1.0cm} (b) 
\includegraphics[scale=0.45,angle=0]{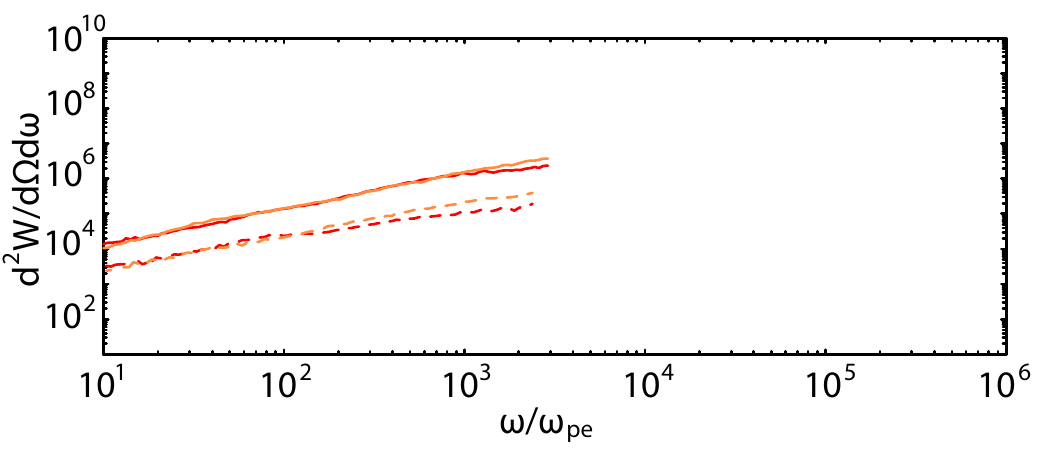}
\includegraphics[scale=0.45,angle=0]{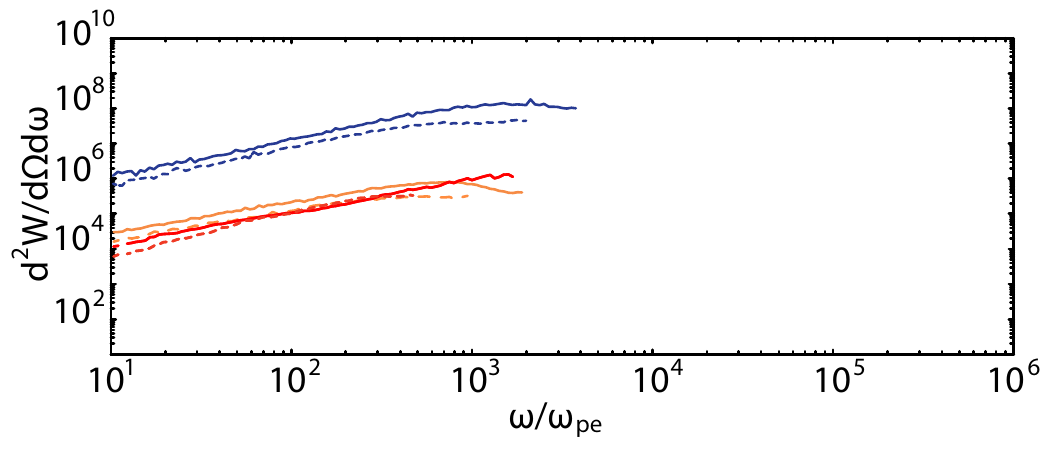}

\hspace*{2.0cm} {\bf e$^{\pm}$ jet with $\mathbf{\Gamma=100}$} \hspace*{1.5cm} (c) \hspace*{3.2cm} {\bf e$^{-}$- i$^{+}$ jet with $\mathbf{\Gamma=100}$} \hspace*{1.0cm} (d) 
\includegraphics[scale=0.45,angle=0]{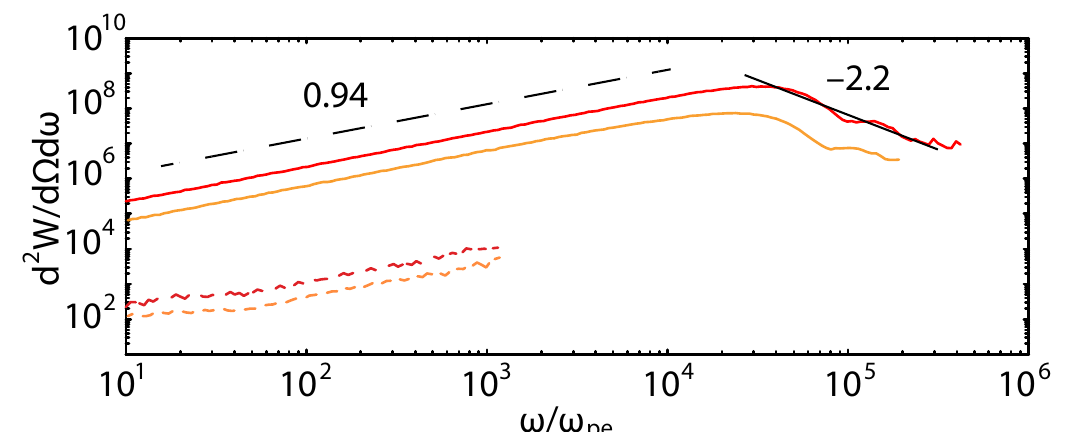}
\includegraphics[scale=0.45,angle=0]{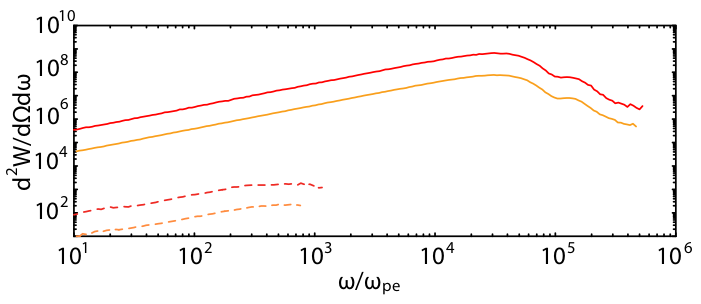}
\end{center}
\caption{Synthetic spectra for e$^{\pm}$ jets (left panels) and for e$^{-}$- i$^{+}$ jets (right panels). Upper panels correspond to jets with a bulk Lorentz factor of $\Gamma=15$ (when selecting electrons with $\gamma>12$, except for the blue lines), while lower panels are for jets with $\Gamma=100$ (when selecting electrons with $\gamma>80$). The continuous line corresponds to the spectra for a head-on emission of jet electrons, whereas the dashed lines represent for 5$^\circ$-off emission of radiation. The red lines show the cases with a stronger amplitude of the initial toroidal magnetic field, $B_{0}=0.5$, whereas the orange lines represent the spectra for $B_{0}=0.1$. In panel (b), the blue lines (solid for the head-on emission and dashed for the 5$^\circ$-off emission) are obtained choosing jet electrons randomly, including all Lorentz factors. In panel (c), the slopes of the power-law segments at lower ($\sim 0.94$) and higher ($\sim -2.2$) frequency are indicated. The slopes for all spectra are listed in Table~\ref{slopes.tab}.}. 
\label{allspectra}
\end{figure*}

\begin{table*}
\begin{center}
\begin{tabular}{l|l|l|l|l|l|l|l}
Panel                    & Emission                & $B_0=0.5$ (red)& $B_0=0.1$ (orange)& $B_0=0.5$ (blue) \\ 
\hline
a) e$^{\pm}$, $\Gamma=15$& head-on (solid)         & $1.00\pm 0.01$ & $0.95\pm 0.02$    &   \\
                         & $5^{\circ}$-off (dashed)& $1.05\pm 0.01$ & $0.93\pm 0.01$    &    \\
b) e$^{-}$-i$^{+}$, $\Gamma=15$ & head-on (solid)  & $0.97\pm 0.01$ & $0.83\pm 0.01$    & $0.84\pm0.01$  \\
                         & $5^{\circ}$-off (dashed)& $0.98\pm 0.03$ & $0.77\pm 0.04$    & $0.76\pm 0.01$   \\
c) e$^{\pm}$, $\Gamma=100$ & head-on (solid)       & $0.94\pm 0.003$ & $0.98\pm 0.003$   &  \\
                         &                         & $\mathit{-2.22\pm 0.1}$ & $\mathit{-1.55\pm 0.08}$ &\\
                         & $5^{\circ}$-off (dashed)& $0.84\pm 0.02$ & $0.90\pm 0.01$    &   \\ 
d) e$^{-}$-i$^{+}$, $\Gamma=100$ & head-on (solid) & $0.87\pm 0.01$ & $0.91\pm 0.01$    &    \\
                         &                         & $\mathit{-2.26\pm 0.10}$ & $\mathit{-2.46\pm 0.13}$ &\\
                         & $5^{\circ}$-off (dashed)& $0.91\pm 0.01$ & $0.92\pm 0.01$    & \\     
\end{tabular}
\caption{Slopes of the spectra presented in Fig. \ref{allspectra}. The four values in italic are the slopes of the decreasing legs of the solid curves (head-on) in panels c) and d).} \label{slopes.tab}
\end{center}
\end{table*}

Depending on the setup of the plasma species (e$^{\pm}$ or e$^{-}$- i$^{+}$) in the jet, we select
electrons with a Lorentz factor $\gamma > 12$ for a jet with $\Gamma=15$ and with $\gamma > 80$ for a jet with $\Gamma=100$, and starting from $t \gtrsim 700 \omega_{pe}^{-1}$ we trace them for 5,000 steps
with a time-step $\Delta t_{\rm s} = 0.005\, \omega_{\rm pe}^{-1}$, thus the jet head moves over 25 grid cells, within a time interval of $t_{\rm s} = 25\, \omega_{\rm pe}^{-1}$. This time-step used for calculating spectra is 200 times smaller than that utilized in the case of the main PIC plasma simulations, $\Delta t = 0.1\, \omega_{\rm pe}^{-1}$, as it is also set in the simulations performed by \citet{meli23}. When tracing the particles, we iterate 10 times over calculating the positions and the velocities of the particles. The Nyquist frequency 
is at $\omega_{\rm N} = 1/(2*0.005*0.1) \, \omega_{\rm pe} = 10^3 \, \omega_{\rm pe}$, where the factor 0.1 in denominator accounts for the fact that we iterate 10 times over when calculating the positions and the velocities of the particles, for each time step of the simulations while performing spectra calculations, which is $\Delta t_{\rm s} = 0.005\, \omega_{\rm pe}^{-1}$. The frequency resolution is $\Delta\omega = 1/t_{\rm s}= 0.04\,\omega_{\rm pe}$, which is small enough to provide smooth spectra.

First, we calculate the radiation spectra for an initial toroidal magnetic field of moderate strength, $B_{0}=0.5$ (see Eqs.~\ref{B1}-\ref{B2}), which corresponds to a plasma magnetization of $\sigma = 1.73\times 10^{-2}$. This setup of the initial toroidal magnetic field is also the case for the studies of plasma instabilities and particle acceleration by \citet{meli23}. 

Here, we analyze the impact of the strength of the initial helical magnetic field on the emission of radiation from the jet electrons for the two types of plasma used to model the jet. In Fig.~\ref{allspectra}, we represent the spectra of the radiated power $P(\omega) = d^2W/d\Omega d\omega$ as a function of the emitted frequencies, $\omega/\omega_{\rm pe}$, for a moderate, $B_{0}=0.5$, initial toroidal magnetic field (red lines) and for a weaker field of $B_{0}=0.1$ (orange lines), for two viewing angles: head-on emission of jet electrons (solid lines) and for 5$^\circ$-off emission (dashed lines). 

For an $e^{\pm}$ jet with $\Gamma=15$ (Fig.~\ref{allspectra}(a)), we cut the spectra at about $10^3\,\omega/\omega_{\rm pe}$, slightly beyond the Nyquist frequency. However, in the case of $e^{-}- i^{+}$ jets with the same $\Gamma=15$ (Fig.~\ref{allspectra}(b)), the spectra show a peak in frequency, even at a higher frequency ($10^4\,\omega/\omega_{\rm pe}$), the location of which depends on how the jet electrons are accelerated by the magnetic (and electric) fields modified from their initial setup by kinetic instabilities. This fact is reflected through the color maps of $B_{y}$, where in Fig~\ref{BFM}(b) for $B_0=0.5$, the amplified magnetic field ($B_{y}^{\max}=\pm 4.14$) is stronger than for the case of $B_0=0.1$ ($B_{y}^{\max}=\pm 2.178$ in Fig~\ref{BFW}(b)), and therefore the jet electrons are much more accelerated in the former case. (We can also observe the difference in particle acceleration by comparing Fig~\ref{LorM}(b) and Fig~\ref{LorW}(b).) In the linear regime of the plasma, the electrons are instantaneously accelerated and decelerated in the highly complicated electric and magnetic fields generated by kinetic instabilities (WI, kKHI, and MI). The WI grows first, then the kKHI and the MI, and later on the MI becomes dominant over the kKHI. In the nonlinear regime, strong electromagnetic fields are generated that lead to particle acceleration. In the turbulent regime, quasi-steady parallel electric fields accelerate the particles further, besides magnetic reconnection might play a role in particle acceleration. Nevertheless, this mechanism of acceleration by quasi-steady parallel electric fields will be addressed in detail in future work.

We have also performed spectra calculation for positrons in the case of an $e^{\pm}$ plasma jet (not shown here). Their spectral characteristics are similar to those for radiation emitted by electrons. Therefore, a factor of two should be added to the emission power to account for both electron and positron emission.

For a jet with $\Gamma=100$ (Figs.~\ref{allspectra}(c,d)), the peak frequency of the spectra is shifted to higher frequencies ($\sim 3\times10^4\,\omega/\omega_{\rm pe}$) than in the cases of $\Gamma=15$, for both plasma compositions. Since in this case the bulk Lorentz factor is $\sim 6.7$ higher, the jet electrons are swerved, producing spectra with higher frequency beyond the Nyquist frequency.
We note that the emission powers for a 5$^\circ$ angle (dashed lines) is by two order of magnitude weaker than that for head-on emission, for a jet with $\Gamma=100$ (Figs.~\ref{allspectra}(c,d)). For $\Gamma=15$ (Figs.~\ref{allspectra}(a,b)), at the two observing angles the spectra are not that much separated. Moreover, these spectra are questionable beyond the Nyquist frequency.

The radiation power shows similar values in both cases of $e^{\pm}$ and $e^{-}- i^{+}$ jets with $\Gamma=100$ (Fig.~\ref{allspectra}(c,d)) since there is not too much difference between the particle species (positrons versus ions with a mass ratio of 4). For a head-on emission from a $e^{\pm}$ jet with $\Gamma=100$ (red lines in Fig.~\ref{allspectra}(c)), the observed difference in the amplitude of the radiation power results from different values of the generated magnetic field, the maximum of which is $B_{y}^{\max}=\pm 1.231$ (Fig.~\ref{BFM}(c)) versus $B_{y}^{\max}=\pm 0.741$ (Fig.~\ref{BFW}(c)). For a jet with $\Gamma=15$, the spectra for head-on and 5$^\circ$-off emission have similar powers, more specifically in the case for an e$^{-}$- i$^{+}$ plasma, whereas for a jet with $\Gamma=100$, the emission power is more than two orders of magnitude stronger for head-on emission than for 5$^\circ$-off emission. This discrepancy arises from the fact that, in the case of a jet with $\Gamma=100$, the plasma instability excited in the transversal plane of the jet is very weak, therefore the electrons are less accelerated along the 5$^\circ$-off direction and, as a consequence, the emitted power is very weak. 

In Fig.~\ref{allspectra}(b), we represent with blue lines the spectra for head on (solid) and 5$^\circ$-off (dashed) emission in the case of an $e^{-}- i^{+}$ jet with $\Gamma=15$ and $B_{0}=0.5$, when we do not impose a lower limit for the Lorentz factor of the electrons that we trace, thus the spectra are obtained by choosing jet electrons randomly including all values of their Lorentz factors. In these cases, both spectra for head-on and 5$^\circ$-off emission present a peak, the amplitude of which is higher than the maximum amplitude that is observed in the spectra when we select electrons with $\gamma > 12$, as the number of the particles that we trace in the former case are higher than that in the latter case.

To calculate the spectra in blue color in Fig.~\ref{allspectra}(b), we pick up randomly from all jet electrons in the PIC simulation, with all possible values for the Lorentz factors of the electrons, only about 8000 electrons (that is because we set the number of the electrons that we can select from a square column, $(14\times14) \Delta^2$, along the jet to 50). {\it A random selection of particles can be regarded as being representative for a system}. In the same Fig. 8b, but in red color, we impose a condition on the Lorentz factor of the jet electrons and select from those $\sim 8000$ jet electrons only the electrons that have $\gamma > 12$. Nevertheless, both kinds of spectra, in blue and in red, show a peak frequency and a power-low slope at low frequencies, typical of a nonthermal radiation. This is also the case for all spectra depicted in Fig. 8. We mention that for a jet with $\Gamma=100$, we select this time electrons with $\gamma > 80$ from those $\sim 8000$ jet electrons.

In order to determine the slopes of all the spectra presented in Fig. \ref{allspectra}, we performed a fit to power laws to each individual line (red, orange, blue; solid, dashed) based on the nonlinear least-squares (NLLS) Marquardt-Levenberg algorithm \cite{levenberg_1944,marquardt_1963}. The resulting slopes, which are the exponents of the power laws, are summarized in Table \ref{slopes.tab}; note that for $e^{\pm}$ and $e^{-}- i^{+}$ jets with $\Gamma=100$, the head-on spectra have an increasing and decreasing leg. For these cases, we have fitted both of them individually and state them separately in Table \ref{slopes.tab}. In the low-frequency region of the spectra, we find that the power-law segment $\sim (\omega/\omega_{\rm pe})^{\alpha}$ has a slope $\alpha \sim 1$ for most cases, exception being for the $e^{-}- i^{+}$ jets with $\Gamma=15$ for two simulation setups (i) $B_0 = 0.1$ and $\gamma > 12$ and (ii) $B_0 = 0.5$ and $\gamma$ randomly selected from all values, where the slope has a lower value, but still larger than that for the synchrotron radiation. The slopes for the $e^{-}- i^{+}$ jets with $\Gamma=15$ and  $B_0 = 0.5$ decreases from $\sim 1$ to $\sim 0.8$ when randomly selecting electrons with all sorts of values of their Lorentz factor for both head-on and 5$^\circ$-off emission. For jets with $\Gamma=100$ and for frequencies above the peak frequency, the spectra follows a power-law $\sim (\omega/\omega_{\rm pe})^{-\beta}$, where $\beta \sim 2.2$.

Our results indicate that the obtained spectra show jitter-like radiation, which, at low frequencies, has a steeper slope ($\sim  0.94$) than the classical synchrotron radiation ($1/3$). As we expected, the peak frequency and the intensity of radiation become higher in the case of a jet with $\Gamma= 100$ (Figs.~\ref{allspectra}(c,d)) than in the case of a jet with $\Gamma= 15$ (Figs.~\ref{allspectra}(a,b)), for the same criteria on tracing jet electrons (red and orange lines).Nevertheless, some differences in the slopes of the spectra can be seen in Table~\ref{slopes.tab} and the intensity of the peak radiation increases by one order of magnitude as the strength of the amplitude of the initial toroidal magnetic field increases by a factor of 5 in the case of a $\Gamma = 100$ jet.  Nevertheless, in future work, we will continue to investigate relativistic jets with stronger initial magnetic fields and with larger jet radius and a non-top-hat jet density profile, which will allow us to accommodate the real value for the ion-to-electron mass ratio. This kind of simulation might excite kinetic kink instability as seen in MHD simulations and jet particles might be accelerated to higher energy than those of present simulations.
 
Our results show steeper spectra than those in the simulations performed by \citet{sironi09}, where the low-frequency part of the spectrum scales as $\omega^{1/3}$. 
The difference can arise from the fact that \citet{sironi09} calculate the spectra from particles accelerated in relativistic collisionless shocks triggered by reflecting an incoming cold upstream flow, where the magnetic fields are generated by the WI. 
Instead, in the current work the spectra are calculated for the electrons selected directly from simulations, whose initial energy distribution can differ. We have calculated the spectra self-consistently, as the magnetic fields and the accelerated particles, key ingredients
for the radiation, are produced as part of the plasma evolution. Therefore, our approach looks to be more natural.

\section{Discussions and conclusions}\label{conclusions}

Using self-consistent, 3D PIC calculations, we have obtained the first synthetic spectra of jitter-like radiation emitted by electrons in relativistic jets containing an initial toroidal magnetic field.  The jitter radiation covers the regime where the magnetic field is inhomogeneous on scales smaller than the Larmor radius and the transverse deflections of the electrons in these fields are much smaller than the relativistic beaming angle. The jitter radiation theory was based on the small-scale nature of the magnetic fields generated by the two-stream instability, where in the one-dimensional analytical approach it was found that the low frequency slope of the spectrum could be steeper than the 1/3 slope of synchrotron radiation \citep{medvedev00}. In our simulations, the jet particles are accelerated along and, to some degree, perpendicularly to the jet direction of propagation (see Figs. 10 (e, f) in \citet{meli23}, for a $\Gamma=15$ jet), but jet electrons propagate rather straight, which is also the case of a $\Gamma=100$ jet (see Figs.\ref{LorM}-\ref{LorW}), therefore the jet electrons radiate jitter-like radiation. Furthermore, the slope of the jitter radiation at low frequencies is steeper $\sim 1$ than the corresponding slope of the classical synchrotron radiation $\sim 1/3$.

We have injected a relativistic jet containing a toroidal magnetic field into an ambient plasma at rest and calculate spectra for two values of the bulk Lorentz factor: $\Gamma=15$ and $\Gamma=100$. (We have used the PIC code in the work by \citep{meli23} and added subroutines for spectra calculation.) These values are typical for jets in AGN and GRB objects, respectively. Furthermore, we set up two values for the amplitude of the initially applied toroidal magnetic field ($B_0=0.5$ and $B_0=0.1$), in order to analyze the impact of the strength of the initial magnetic field on the characteristics of the calculated spectra.

To estimate the characteristic length-scale of the turbulence, we have used the FFT to convert the amplitude of the $B_{y}$ wave components of the magnetic field from the spatial domain to the wave vector domain. We have found that the strongest mode corresponds to a wavelength of $\lambda_{\perp} = (24 \pm 1)\,\Delta = (2.4 \pm 0.1)\,\lambda_{\rm e}$, where the electron plasma skin depth is $\lambda_{\rm e} = 10\,\Delta$, in our simulations. For comparison, the jitter radiation is expected to emerge from relativistic particles traveling through small-scale, turbulent magnetic fields that are coherent on the scale of an electron plasma skin depth, where in the work by \citet{medvedev00} the origin of the magnetic field comes from a relativistic version of the well-known two-stream WI in a plasma.

We have calculated synthetic spectra directly from self-consistent simulations in which growing kinetic instabilities (WI, kKHI, and MI) are developed simultaneously into a highly nonlinear regime of plasma for two types of particle species of the jet plasma (pair and electron-ion). The strength and structure of growing instabilities depend on many factors, such as the amplitude of the initial toroidal magnetic field, the bulk Lorentz factor of the jets, and the type of the jet plasma (see Figs.~\ref{BFM} and \ref{BFW}).

In recent years, several methods have been developed for calculating synthetic spectra using PIC simulations \cite[e.g.,][]{reville10,kagan16,spisak20,zhang23}. 
Nevertheless, the closest results with which our synthetic spectra can be compered are in the papers by \citet{sironi09} and, more recently, by \citet{zhang23}. 

On the one hand, in the paper by \citet{sironi09}, in the 3D case, the low-frequency part of the spectrum scales as $\omega^{1/3}$, which can be attributed to synchrotron radiation. In 2D, they calculate the spectra with artificially reduced intensity of the magnetic field and obtain a flatter slope as for a jitter-like radiation. Our results show jitter-like spectra with a steep slope, $\omega^{0.94}$, generated from jet electrons accelerated in a turbulent magnetic field, which resulted from growing instabilities (WI, kKHI, and MI), in the presence of an initial toroidal magnetic field, into the nonlinear regime. 

On the other hand, \citet{zhang23} perform combined 2D PIC and polarized radiative transfer simulations to study synchrotron emission from magnetic turbulence in the blazar emission region. The calculated spectra are shown in the upper panel of Fig. 3, where a flatter slope can be observed when the turbulent magnetic field is weaker. As their simulations are only 2D, this may be also a source of important differences.

To explore applications of our results, and to offer another perspective at understanding the 
role of the physical conditions of jet plasma on the electron acceleration and emission 
processes, we qualitatively compare our simulations to observations of relativistic jets. PIC calculations are carried out in dimensionless grid units which must be scaled (via scaling factors) into physical units (i.e., cgs) \cite[e.g.,][]{macdonald21}.

To rescale the radiation spectra obtained from simulations, we have to take into account the effects induced by (i) simulation time-scale (i.e., $\omega_{\rm pe}/\omega_{\rm pe}^{\rm sim}$) and (ii) relativistic Doppler shift ($\sim 2~\Gamma$), as the jet moves with a bulk Lorentz factor $\Gamma$ with respect to an observer at rest \citep{hededal05}.

Now, to rescale the time-scale of the simulations to the real space we divide all frequencies with the plasma frequency calculated in the simulation box, $\omega_{\rm pe}^{\rm sim}$, and multiplying them with the real $\omega_{\rm pe}$. From the simulation code, the electron plasma rest-frame frequency is calculated as $\omega_{\rm pe}^{\rm sim} = \sqrt{q_{\rm e}^2 a_{\rm dens}/m_{\rm e}}$, where $q_{\rm e} = -0.01$, $a_{\rm dens} = 12.0$, and $m_{\rm e}=0.12$. 
This means, $\omega_{\rm pe}^{\rm sim} = 0.1 \, \Delta_{\rm t}^{-1}$, where $\Delta_{\rm t}$ is the simulation unit time, which in our computations is set to unity. 

The electron plasma rest-frame frequency in the jet is (in cgs units): $\omega_{\rm pe}=(4\pi e^2 n_{\rm e}/m_{\rm e} )^{1/2} =  5.64\times 10^4 \, (n_{\rm e}/\rm{cm}^{-3})^{1/2}$, where $n_{\rm e}$, $e$, and $m_{\rm e}$ denote the number density, charge, and mass of the electrons within the plasma.

Next, we scale the PIC value of the number density to a fiducial value, which is estimated from observations,  $n_{\rm e}= n_{\rm e}^{\rm obs}$. Such a value is not universal, being rather tailored to each jet in AGN or GRB objects. Here, we consider the case of the M87 jet, where the electron number density of the jet is $n_{\rm e}^{\rm M87}\lesssim 10^{4}$ cm$^{-3}$ \citep{kawashima22}, since the synchrotron flux from the jet should be less than the observed flux in M87 by Event Horizon Telescope observations \citep{eht19}. 

Therefore, $\omega_{\rm pe}  \lesssim 5.64\times10^6$ s$^{-1}$ (in the rest-frame of the jet) and, the frequency range of spectra is $5.64\times10^7$ Hz -- $5.64\times10^{10}$ Hz for relativistic jets with a similar jet density like M87 jet. Thus, the frequency axis of the spectra in Fig.~\ref{allspectra} emitted by electrons in a jet (with $n_{\rm e} \sim n_{\rm e}^{\rm M87}$) that propagates with a bulk Lorentz factor of $\Gamma = 15$ and $\Gamma = 100$ should be shifted to higher frequencies by $\sim 5.64\times10^7 \cdot 2 \cdot 15$ Hz $\simeq 1.7\times10^{9}$ Hz and $\sim 5.64\times10^7 \cdot 2 \cdot 100$ Hz $\simeq 1.1\times10^{10}$ Hz, respectively. After shifting the frequencies in Fig.\ref{allspectra}, the observed frequencies are already in the X-ray domain ($\sim 10^{16}-10^{19}$ Hz). Furthermore, the value $n_{\rm e}^{\rm M87}\lesssim 10^{4}$ cm$^{-3}$ is rather an estimation of the electron number density at the base of the jet. We expect to have a few order of magnitude lower values for $n_{\rm e}$ far from the black hole. In this case, the electron plasma rest-frame frequency, $\omega_{\rm pe}$, can have lower values, thus the frequency range of the emitted spectra in Fig.~\ref{allspectra} and the observed spectra can be shifted to higher frequencies than those calculated above.

To estimate the length scale of the PIC jet, we use the expression for  electron plasma rest-frame frequency above, plus including the relativistic Lorentz factor of the jet $\Gamma$, in a similar way as in Eqs. (1)-(7) of \citet{macdonald21}, by taking $n_{\rm e} = n_{\rm e}^{\rm M87}\lesssim 10^{4}$ cm$^{-3}$. The length scale, $\sim c/[5.64\times 10^4 \, \Gamma^{-1/2}(n_{\rm e}/\rm{cm}^{-3})^{1/2}]$, becomes $\sim  2 \times 10^4$ cm and $\sim 5.3 \times 10^4$ cm for $\Gamma = 15$ and $\Gamma = 100$, respectively. This scaling implies that the jet length of our PIC simulations (i.e., 1285 cells in length) corresponds to a physical size of a few hundreds of km. Such a PIC jet length is much less than the length of an AGN jet \citep{walg13}, but similar to that of a jet in an X-ray binary, where the derived bulk Lorentz factors of the jets, which are in most cases lower limits, are found to be large, with a mean $ > 10$, comparable to those estimated for AGN \citep{miller06}. However, these estimates of the physical lengths of the PIC jet represent lower limits, as we use for $n_{\rm e}$ a value which is rather at the base of the jet. Since far from the black hole, we expect to have a few order of magnitude lower values for $n_{\rm e}$ than $10^{4}$ cm$^{-3}$, the physical size of the PIC jet can increase at least two orders of magnitude if $n_{\rm e} \sim 1$ cm$^{-3}$, but larger grid simulations are needed.

Although MHD (or MHD-PIC) simulations can provide a length scale of the jet similar to that of an AGN or GRB jet (as the MHD method is scale free), we cannot perform in the current paper a direct comparison of our results with those of current MHD simulations \cite[e.g.,][]{medina21,medina23}. A more detailed quantitative analysis, including the computation of the acceleration rate and the properties of the reconnection layers driven by the instabilities and turbulence will be described in future work, for an appropriate comparison. In this paper, we focus on relevant kinetic-scale physics within relativistic jet plasmas.

In our simulations, we consider two types of plasma composition for both the jet and the ambient medium. A pair plasma and an electron-ion plasma, where the mass ratio is $m_{\rm i}/m_{\rm e} = 4$. We select this value for the ion-to-electron mass ratio because it results in a significantly weaker growth of the MI compared to higher mass ratios.
(For more details on the selection $m_{\rm i}/m_{\rm e} = 4$, see the discussion in \citet{meli23}.) Although, the real mass ratio for a proton is $m_{\rm p}/m_{\rm e} = 1836$, we do not make a comparison between a pair and a proton-electron plasma jet, from the observational point of view, instead, we are interested in discerning the trends in the radiation spectra as we increase the mass ratio from 1 to 4. Calculations of radiation spectra for a proton-electron plasma with a mass ratio that more closely reflects a realistic value will be included in further work, where larger simulation grids are needed to accommodate the growth of MI. 

This is the first in a series of papers that aim to calculate synthetic spectra from PIC simulations of relativistic plasma jets, where the jets are injected into a very large simulation grid, where no periodic boundary conditions are applied in the direction of jet propagation. The present simulations are designed to trace electrons with a high Lorentz factor from a region located at the head of the jet, for a relatively short time (due to the very high computational costs). Different selection criteria for tracing the electrons and the use of the FFT analysis of turbulence for finding the strongest modes to localize the region from where to trace electrons will be addressed in the following papers.

Further work should also exploit larger scale simulations, including cooling terms in the equation for the radiation power and an improved initial set-up (e.g., Gaussian jet density profile instead of the top-hat one, which is utilized in the work presented here, different plasma composition for the jet and for the ambient, more realistic mass ratios for ions or stronger initial magnetic fields), just to account for a more realistic description of the emission from relativistic jets. Larger simulation systems would also allow us to verify whether the magnetic field modified globally by kinetic instabilities (WI, kKHI, and MI) in the presence of an initial toroidal magnetic field can avoid dissipation and survive beyond a few hundred electron skin depths and to unveil the occurrence of plasma shocks by the presence of sharp variations in the magnetic field profiles. Thus, we should be able to calculate synthetic spectra in more specific and realistic jet conditions based on observations. Certainly, large scale MHD simulations with test particles can complement PIC simulations, but they cannot address the kinetic effects as PIC simulations do.

\section*{Acknowledgments}
The authors would like to thank the collaborators Jacek Niemiec and Martin Pohl for valuable discussions during the development of this work. We are grateful to the anonymous reviewer for their thoughtful questions and suggestions that subsequently improved the quality of the paper. 
The simulations presented in this report  have been performed on the Frontera supercomputer at the Texas Advanced Computing Center under the AST 23035 award: PIC Simulations of Relativistic Jets with Toroidal Magnetic Fields (PI: Athina Meli) through the NSF grant No 2302075; and the  AST21038 award: Computational Study of Astrophysical Plasmas; through the NASA grant: Nature Of Hard X-rays From A TeV-detected RadioGalaxy (PI: Ka Wah Wong at SUNY Brockport) issued by the  NuSTAR Guest Observer Cycle 6 2019; and using the Pleiades facilities at the NASA Advanced Supercomputing (NAS: s2004 and s2349), which is supported by the NSF; as well as Ares supercomputer at Cyfronet AGH (PI: Oleh Kobzar) through the grant PLG/2024/017211. I.D. acknowledges support from the Romanian Ministry of Research, Innovation and Digitalization under the Romanian National Core Program LAPLAS VII - contract no. 30N/2023. K.N. and A.M. acknowledge support from the  NSF Excellence in Research Award No (FAIN): 2302075. O.K. is supported by the Polish NSC (grant 2016/22/E/ST9/00061). C.K. has received funding from the Independent Research Fund Denmark (grant 1054-00104). Y.M. is supported by the ERC Synergy Grant ``BlackHoleCam: Imaging the Event Horizon of Black Holes'' (Grant No. 610058). JLG acknowledges the support of the Spanish Spanish Ministerio de Ciencia, Innovaci\'{o}n y Universidades (grants PID2019-108995GB-C21 and PID2022-140888NB-C21) and the State Agency for Research of the Spanish MCIU through the Center of Excellence Severo Ochoa award for the Instituto de Astrof\'{\i}sica de Andaluc\'{\i}a (CEX2021-001131-S).


\section*{Data Availability}

The data underlying this article will be shared on reasonable request to the corresponding author.





\bsp	
\label{lastpage}
\end{document}